\documentclass{article}
\usepackage{latexsym}
\usepackage{amsfonts}

\usepackage{epsfig}
\usepackage{graphicx}

\def\C{\mathbf{C}}

\newcommand{\BP}{{\mathbb P}}
\def\Z{\mathbf{Z}}

\newcommand{\R}{{\mathbb R}}

\def\a{\alpha}
\def\b{\beta}

\def\ve{\varepsilon}
\def\e{\epsilon}

\def\s{\sigma}
\def\rhi{\varphi}

\def\A{{\mathcal A}}
\def\CC{{\mathcal C}}

\renewcommand{\(}{\left(}
\renewcommand{\)}{\right)}

\def\oinn#1#2{\left\langle#2\,\left\vert\,#1\right.\right\rangle} 
\def\inn#1#2{\left\langle#1\,\left\vert\,#2\right.\right\rangle}
\def\INN{\langle\cdot\,\vert\,\cdot\rangle}

\def\diag{\mathop{\hbox{\rm diag}}\nolimits}

\newtheorem{theorem}{Theorem}[section]
\newtheorem{proposition}[theorem]{Proposition}
\newtheorem{cor}[theorem]{Corollary}

\newtheorem{lemma}[theorem]{Lemma}
\newtheorem{definition}[theorem]{Definition}

\def \qed{{\hfill $\Box$}}

\newenvironment{proof}
  {{\sl Proof}}
  {\par\smallskip}

\newenvironment{equation*}
  {$$}
  {$$}

\newenvironment{eqn*}[1][1.5]
  {$$\renewcommand{\arraystretch}{#1}
      \begin{array}{rcl}}
      {\end{array}$$}

\newenvironment{eqn}[2][1.5]
  {\begin{equation}\label{#2}
   \renewcommand{\arraystretch}{#1}
   \begin{array}{rcl}}
  {\end{array}\end{equation}}

\renewenvironment{matrix}[1]{\left(\begin{array}{#1}}{\end{array}\right)}

\catcode`\@=11
\def\@nodimen#1{\expandafter\@@nodimen\the#1}%
{\catcode`\p=12\catcode`\t=12\gdef\@@nodimen#1pt{#1}}%
\def\crossed#1{{\setbox0=\hbox{$#1$}%
    \dimen@=\wd0\dimen@i=\ht0\dimen@ii=\dp0
    \advance\dimen@ by 0pt
    \advance\dimen@i by 2pt
    \advance\dimen@ii by 2pt
    \hbox{\special{ps:
}$#1$}}}%
\catcode`\@=12

\def\Type#1#2{$\mbox{\em Type #1 }_{|_{#2}}$}
\def\dint{\int\!\!\!\int}
\def\exi#1#2{e^{\xi{(#1,#2)}}}
\def\emi#1#2{e^{-\xi{(#1,#2)}}}
\def\ebi#1#2{e^{\pm\xi{(#1,#2)}}}

\def\zs{\left[z^\mi\right]}
\def\ds{\displaystyle}
\def\comment#1{}  
\def\II{\sqrt{-1}}
\def\mi{{-1}} 

\def\p{\partial}
\def\pp#1#2{\frac{\p #1}{\p #2}}
\def\transp#1{{#1^\top}} 

\def\iy{\infty}
\def\set#1{\left\{#1\right\}}

\def\bddots{\mathinner{\mkern1mu\raise1pt\vbox{\kern7pt\hbox{.}}\mkern2mu
             \raise4pt\hbox{.}\mkern2mu\raise7pt\hbox{.}\mkern1mu}}

\def\?{(?)\marginpar{|?}}

\def\P#1{P^{(#1)}_{mn}}
\def\PS#1{P^{*(#1)}_{nm}}
\def\PSS#1{P^{*(#1)}_{n^*m^*}}
\def\Q#1{Q^{(#1)}_{mn}}
\def\QS#1{Q^{*(#1)}_{nm}}

\begin{document}
\nocite{*}
\centerline{\LARGE Moment matrices and multi-component KP,}
\centerline{\LARGE  with applications to random matrix theory}
\bigskip
\centerline{Mark Adler\footnote{Department  of Mathematics, Brandeis University, Waltham, Mass 02454, USA,
adler@brandeis.edu. The support of a National Science Foundation grant \# DMS-04-06287 is gratefully acknowledged},
Pierre van Moerbeke\footnote{D\'epartement de Math\'ematiques, Universit\'e Catholique de Louvain, 1348
Louvain-la-Neuve, Belgium and Brandeis University, Waltham, Mass 02454, USA, vanmoerbeke@math.ucl.ac.be. The
support of a National Science Foundation grant \# DMS-04-06287, a European Science Foundation grant (MISGAM), a
Marie Curie Grant (ENIGMA), Nato, FNRS and Francqui Foundation grants is gratefully acknowledged.} and Pol
Vanhaecke\footnote{The support of a European Science Foundation grant (MISGAM) and a Marie Curie Grant (ENIGMA) is
gratefully acknowledged.}}
%
%
%
%
\tableofcontents

\section{Introduction}
Random matrix theory has led to the discovery of novel matrix models and novel statistical distributions, which are
defined by means of Fredholm determinants and which, in many cases, satisfy nonlinear ordinary or partial
differential equations. A crucial observation is that these matrix integrals, upon appropriate deformation by means
of exponentials containing one or several series of time parameters, satisfy (i) integrable equations and (ii)
Virasoro constraints with respect to these time parameters. Most of the time, such matrix integrals can be written
--- by expressing the integrand in ``polar coordinates" --- as a multiple integral, which then can be expressed in
terms of the determinant of a moment matrix; this may be a moment matrix with regard to one or several weights. The
extra time parameters are added in such a way that each weight has its own exponential time deformation.

\smallskip

The main point is to show that this determinant satisfies (i) and (ii). These features turn out to be extremely
robust! The purpose of the present paper is to show point (i) in great generality, which is the determinant of
moment matrices associated with one or several weights and defined on various different domains, satisfies the {\em
multi-component KP hierarchy} with regard to the time parameters. This is a very general class of integrable
equations.

\smallskip

This determinant will turn out to be the $\tau$-function of this integrable hierarchy; this $\tau$-function with
appropriate shifts of the deformation variables will be expressed in terms of the ``orthogonal polynomials" defined
by the weights and their Cauchy transform. We list below a number of examples having their origin in Hermitian
random matrix theory, in random matrices coupled in a chain, in random permutations and in Dyson Brownian motions
(non-intersecting Brownian motions) on $\R$ leaving from the origin, where some paths are forced to end up at one
point and others at another point, etc\dots These examples will then be discussed in detail in Section
\ref{sec:examples}.

\bigskip
\noindent $\bullet$\ \emph{GUE: orthogonal polynomials.}
\begin{equation*}
    {\frac{1}{n!}\int_{E^n}\Delta^{2}_n(z)\prod^n_{\ell=1} e^{\sum_{k=1}^{\iy}t_kz_\ell^k}\rho(z_\ell)dz_\ell} 
     =\det \Bigl(\int_{\R}z^{i+j}e^{\sum_{k=1}^{\iy}t_kz^k}\rho(z)dz\Bigr)_{0\leq i,j\leq n-1}
\end{equation*}

\medskip
\noindent $\bullet$\ \emph{Coupled random matrices / Dyson Brownian motions: bi-orthogonal \hfill\break \hphantom{a} polynomials.}
\begin{eqnarray*}
  \lefteqn{\frac{1}{n!}\int \!\!\!\!\int_{E^n}\Delta_n(x)\Delta_n(y)
    \prod^n_{\ell=1}e^{\sum^{\iy}_{k=1}(t_kx_\ell^k-s_k y^k_\ell)}\rho(x_\ell,y_\ell)dx_\ell dy_\ell }\hspace*{6cm} \\
  &&\hspace*{-4cm}=\det \left(\int\!\!\!\!\int_{E}x^iy^je^{\sum^{\iy}_{k=1}(t_kx^k-s_ky^k)}\rho(x,y)dxdy
    \right)_{0\leq i,j\leq n-1}
\end{eqnarray*}

\medskip
\noindent $\bullet$\ \emph{Longest increasing subsequences in random permutations: orthogonal\hfill\break\hphantom{a}
    polynomials on $S^1$.}
\begin{eqnarray*}
  \lefteqn{\frac{1}{n!} \int_{(S^1)^{n}}|\Delta_n(z)|^{2}\prod_{\ell=1}^n
  \left(e^{\sum_{k=1}^{\iy}(t_k z_\ell^k-s_kz_\ell^{-k})}\frac{dz_\ell}{2\pi \sqrt{-1} z_\ell}\right)}\hspace*{6cm} \\
 &&\hspace*{-4cm}=\det\(\oint_{S^1}\frac{dz}{2\pi\sqrt{-1}z}z^{i-j}e^{\sum_{k=1}^{\iy}(t_kz^k-s_kz^{-k})}\)_{0\leq i,j\leq n-1}
\end{eqnarray*}

\medskip
\noindent $\bullet$\ \emph{$m_1+m_2$ non-intersecting Brownian motions on $\R$ leaving from $0$
  and\hfill\break\hphantom{a} $m_{\!\!\!\tiny\begin{array}{l}1\\2\end{array}}$ paths forced to end up at $\pm a$:
  multiple orthogonal polynomials on $\R$.}
\begin{eqnarray*}
  \lefteqn{\frac{1}{m_1!m_2!}\int_{E^{m_1+m_2 }}\Delta_{m_1+m_2 }(x,y)}\hspace*{6cm} \\
  \lefteqn{\(\Delta_{m_1}(x)\prod^{m_1}_{\ell=1}e^{-\frac{x_\ell^2}{2}+ax_\ell}e^{\sum_{k=1}^{\iy}(t_k-s_k)x^k_\ell}dx_\ell\)}
               \hspace*{6cm}\\
  \lefteqn{\(\Delta_{m_2}(y)\prod^{m_2}_{\ell=1}e^{-\frac{y_\ell^2}{2}-ay_\ell}e^{\sum_{k=1}^{\iy}(t_k-u_k)y^k_\ell}dy_\ell\)}
               \hspace*{6cm}\\
  &&\hspace*{-6cm}=\det\left(
     \begin{array}{c}
       \left(\displaystyle{\int_{E}}z^{i+j}e^{-\frac{z^2}{2}+az}e^{\sum_1^{\iy}(t_k-s_k)z^k}dz\right)_{\tiny{
          \begin{array}{l}
            0\leq i\leq m_1-1\\
            0\leq j\leq m_1+m_2-1
          \end{array}}}\\
     \left(\displaystyle{\int_{E}}z^{i+j}e^{-\frac{z^2}{2}-az}e^{\sum_1^{\iy}(t_k-u_k)z^k}dz\right)_{\tiny{
          \begin{array}{l}
            0\leq i\leq m_2-1\\
            0\leq j\leq m_1+m_2-1
        \end{array}}}
     \end{array}
     \right)
\end{eqnarray*}

\medskip
\noindent $\bullet$\ \emph{$\sum_{\a=1}^q m_{\a}=\sum_{\b=1}^p n_{\beta}$ non-intersecting Brownian motions on
$\R$, with \hfill\break \hphantom{a} $m_{\a}$ paths starting at $a_{\a}\in\R$ and $n_{\beta}$ paths forced to end
up at $b_{\beta}$: \hfill\break \hphantom{a} mixed multiple orthogonal polynomials (mixed mops) on $\R$.}

\bigskip

\noindent{\bf A moment matrix for several weights}: Define two sets of weights
$$
 \psi_1(x),\ldots,\psi_q(x)~~~\mbox{and}~~~\varphi_1(y),\ldots,\varphi_p(y),~~~~~~~~\mbox{with}~x,y\in \R,
$$
and deformed weights depending on {\it time} parameters $s_\a=(s_{\a 1},s_{\a2},\ldots)$ ($1\leq \a\leq q$) and
$t_\b=(t_{\b1},t_{\b2},\ldots)$ ($1\leq \b\leq p$), denoted by
$$
  \psi_\a^{-s}(x):=\psi_\a(x)e^{-\sum_{k=1}^\iy s_{\a k}x^k}
   ~~~~~\mbox{and}~~~~~
  \varphi_\b^{t}(y):=\varphi_\b(y)e^{\sum_{k=1}^\iy t_{\b k}y^k}.
$$
That is, each weight goes with its own set of times. For each set of positive integers\footnote{$\vert m
\vert=\sum_{\a=1}^{q}m_\a$ and $\vert n\vert=\sum_{\b=1}^{p}n_\b$.}
$$
  m=(m_1,\ldots,m_q),n=(n_1,\ldots,n_p)\mbox{~ with~}\vert m\vert=\vert n\vert,
$$
consider the determinant of a moment matrix $T_{mn}$,
composed of blocks and of size $\vert m\vert=\vert
n\vert$, with regard to a (not necessarily symmetric)
inner product $\INN$
\begin{eqnarray}\label{intro1}
  \lefteqn{\tau_{mn}(s_1,\ldots,s_q;t_1,\ldots,t_p)}
  \hskip1cm\nonumber\\ \nonumber\\ \nonumber
  &&  \hspace*{-1cm}
  := \det T_{mn}
  \nonumber\\ \nonumber\\
   && \hspace*{-1cm}:=  \det\begin{matrix}{ccc}
    T_{mn}^{11}&\dots&T_{mn}^{1p}\\
    \vdots&&\vdots\\
    T_{mn}^{q1}&\dots&T_{mn}^{qp}
  \end{matrix}  \nonumber\\
  &&  \hspace*{-1cm}
  := \det\left(\!\!\!\!\begin{array}{ccc}
  \Bigl(\!
  \inn{x^i\psi^{-s}_1(x)} {y^j\varphi^{t}_1(y) }
  \Bigr)_{0\leq i<m_1\atop{0\leq j<n_1}}&\ldots&\left(
  \inn{x^i\psi^{-s}_1(x)} {y^j\varphi^{t}_p(y) }
  \!\right)_{0\leq i<m_1\atop{0\leq j<n_p}}\\
  & & \\
  \vdots& &\vdots\\
  & &\\
  \left(\!\inn{x^i\psi^{-s}_q(x)}
  {y^j\varphi^{t}_1(y) } \right)_{0\leq i<m_q\atop{0\leq j<n_1}}&\ldots&\left(\inn{x^i\psi^{-s}_q(x)}
  {y^j\varphi^{t}_p(y)}\! \right)_{0\leq i<m_q\atop{0\leq j<n_p}}
  \end{array}\!\!\!\!\right).\nonumber\\ \label{Tmn_def}
\end{eqnarray}
A typical inner product to keep in mind is
\begin{equation}\label{example}
  \inn{f(x)}{g(y)}=\int\!\!\!\!\int_{\R^2}f(x)g(y)\,d\mu(x,y),
\end{equation}%
where $\mu=\mu(x,y)$ is a fixed measure on $\R^2$, perhaps having support on a line or curve.

\bigskip

\noindent{\bf From moment matrices to polynomials and their Cauchy transforms}:

\smallskip
{\bf I.}
Then, for $1\leq \b,\,\b'\leq p$, the following
expressions are polynomials (with coefficients depending
on $s$ and $t$)\footnote{Introduce the notation
$[\a]:=(\a,\frac{\a^2}{2},\frac{\a^3}{3},\ldots)$ for
$\a\in \C$. Only shifted times will be made explicit in
the $\tau$-functions; i.e., $\tau_{mn}(t_{\ell}-
{\left[z^{-1}\right]})$ means that $\tau_{mn}$ still
depends on all time parameters, but the variable
$t_{\ell}$ only gets shifted. Moreover, here and below
all the expressions $\ve_{\a\beta}(n) $,
$\e_{\a\beta}(n,m)$, etc$\dots$ all equal $\pm 1$ and
will be given later. Throughout the paper, we use the
standard notation
$e_{1}=(1,0,0,\ldots),~e_{2}=(0,1,0,\ldots)
 $.}
\begin{equation}\label{introI}
  \begin{array}{rcl}
    {\displaystyle z^{n_\b}\frac{\tau_{mn}(t_{\b}-{\left[z^{-1}\right]})}
      {\tau_{mn}}}&:=&\Q{\b,\b}(z)=z^{n_{\b}}+\ldots\\
       {\displaystyle \ve_{\b\b'}(n)z^{n_{\b'}-1}\frac{\tau_{m,n+e_\b-e_{\b'}}(t_{\b'}
      -{\left[z^{-1}\right]})}{\tau_{mn}}}
      &:=&\Q{\b,\b'}(z),\qquad  \left\{\begin{array}{l}
         \mbox{of degree}~<n_{\b'}\\
         \mbox{for}~ \b' \neq\b,  \end{array}\right.
    \end{array}
\end{equation}
satisfying, for each $\beta$, the following
orthogonality conditions
%
\begin{equation}\label{perp1}
  \inn{ x^i\psi_{\a}^{-s}(x)}{\sum_{\b'=1}^p \Q{\b,\b'}(y )
  \varphi_{\b'}^{t}(y)}=0 \mbox{~~for~~}\left\{
  \begin{array}{l}
    1\leq\a\leq q \\
    0\leq i\leq m_{\a}-1.
  \end{array}\right.
\end{equation}

\medskip

\noindent
{\bf II.}  In the same way, the following expressions are polynomials (depending on $s$ and $t$)
\begin{equation}\label{introII}
  \e_{\b\a}(n,m)\,z^{ m_\a-1}\,\frac{\tau_{m-e_\a,n-e_\b}(s_\a+[z^{-1}])}{\tau_{mn}}
    = \PS{\b,\a}(z )~~\mbox{of degree}<m_{\a}
\end{equation}
satisfying, for each $\beta$, the orthogonality
relations 
%
\begin{equation}\label{perp2}
   \left\{
   \begin{array}{l}
      \inn{\displaystyle\sum_{\a=1}^q \PS{\b,\a}(x)\psi^{-s}_{\a}
      (x)} {\begin{array}{l} \\ \\   \end{array}y^j\varphi^t_{\b'}(y)}=0
      \mbox{~~for}\left\{
       \begin{array}{l}
         1\leq\b'\leq p,~0\leq j\leq n_{\b'}-1\\
         \mbox{except}~\b'=\b,~j=n_{\b}-1
       \end{array}\right.\\ \\
      \inn{\displaystyle\sum_{\a=1}^q \PS{\b,\a}(x)
      \psi^{-s}_{\a}(x)} {\begin{array}{l} \\ \\   \end{array}y^{n_\b -1}\varphi^t_{\b}(y)}=1.
   \end{array}\right.
\end{equation}

\medskip
\noindent
{\bf III.} The following expressions are Cauchy transforms of the polynomials obtained in II:

\begin{equation}\label{introIII}
    \begin{array}{rcl}
      {\displaystyle z^{-n_\b}
      \frac{\tau_{mn}(t_{\b}+{\left[z^{-1}\right]})}{\tau_{mn}}}&=&
      \inn{{\displaystyle\sum_{\a=1}^q \PS{\b,\a}}(x) \psi^{-s}_{\a}(x)}
      {\displaystyle\frac{\varphi^{t}_{\b}(y)}{z-y}}\nonumber\\
\nonumber\\
       {\displaystyle \ve_{\b\b'}(n)z^{-n_{\b'}-1}
      \frac{\tau_{m,n+e_{\b'}-e_{\b}}(t_{\b'}+{\left[z^{-1}\right]})}{\tau_{mn}}}
      &=&
      \inn{{\displaystyle\sum_{\a=1}^q
      \PS{\b,\a}(x)
      \psi^{-s}_{\a}(x)}}
      {\displaystyle\frac{\varphi^{t}_{\b'}(y)}{z-y}}
     \end{array}
\end{equation}

\medskip
\noindent
{\bf IV.} Similarly, the following expressions are Cauchy transforms of the polynomials obtained in I:
\begin{equation}\label{introIV}
  \e_{\a\b}(m,n)\,z^{-m_\a-1}\,\frac{\tau_{m+e_\a,n+e_\b}(s_\a-[z^{-1}])}{\tau_{mn}}
         =\inn{\frac{\psi^{-s}_{\a}(x)}{z-x}}{\sum_{\b'=1}^p\Q{\b,\b'}(y)\varphi_{\b'}^{t}(y)}.
\end{equation}

\medbreak

\noindent The statements I, II, III and IV summarize sections 1, 2 and 3.
 As will appear in Section 2, the 
  polynomials appearing in ({\bf I}) are called \Type
{II}{\varphi^t_\b} mixed multiple orthogonal
polynomials, whereas those appearing in ({\bf II}) \Type
I{\psi_\a^t} mixed multiple orthogonal polynomials.
These were introduced by E. Daems and A. Kuijlaars
\cite{DK}, in the context of non-intersecting Brownian
motions; they are a generalization of multiple
orthogonal polynomials, where instead of one set of
weights, there are two sets (the classical orthogonal
polynomials correspond to one set with one element).
They were introduced and studied by Aptekarev, Bleher,
Geronimo, Kuijlaars, Van Assche \cite{Van Assche1, Van
Assche2, Van Assche3, BleKui3}. Around the same time,
they were introduced by Adler-van Moerbeke in the
context of band matrices and vertex operator solutions
to the KP hierarchy \cite{AMgop}. In
\cite{BleKui2,BleKui3}, they were used in the context of
non-intersecting Brownian motions and random matrices
with external source.

\bigskip

\noindent{\bf The $(p+q)$-KP hierarchy}:
Define two matrices $W_{mn}(z)$ and $W^*_{mn}(z)$ of
size $p+q$, whose entries are given by ratios of
determinants $\tau_{mn}$ of moment matrices as above,
but with appropriately shifted $t$ and $s$ parameters.
They turn out to be the wave and dual wave matrices for
the {\em $(p+q)$-KP hierarchy}. It is remarkable that,
upon setting all $t$ and $s$ parameters equal to zero,
the matrix $W_{mn}(z)$ below is precisely the {\em
Riemann-Hilbert} matrix characterizing the mixed
multiple orthogonal polynomials! Similarly $W_{mn}^*(z)$
at $t=s=0$ satisfies the Riemann-Hilbert problem
characterizing alternately the ``dual" multiple
orthogonal polynomials or the inverse transpose matrix
of $W_{mn}(z)$ at $t=s=0$. The Riemann-Hilbert matrix
for the multiple-orthogonal polynomials has been defined
in Daems-Kuijlaars \cite{DK}, which is a far
generalization of the Riemann-Hilbert matrix of
Fokas-Its-Kitaev \cite{FIK} and Deift-Zhou \cite{Deift}.
Using identities as in I to IV, the two left blocks of
$W_{mn}$ and the two right blocks of $W^*_{mn}$ are
mixed multiple orthogonal polynomials, and the remaining
blocks are Cauchy transforms of such polynomials; for
explicit expressions, see Section \ref{sec:RH}. The
matrix $W_{mn}(z)$ is defined by

\begin{equation*}
  W_{mn}(z)\diag \left(e^{-\sum_1^{\iy}t_{1k}z^k},\ldots,e^{-\sum_1^{\iy}t_{pk}z^k},
  e^{-\sum_1^{\iy}s_{1k}z^k},\ldots,e^{-\sum_1^{\iy}s_{qk}z^k}\right):=
  \hspace*{3cm} \renewcommand{\arraystretch}{2}
%
\end{equation*}%
\begin{equation}\label{intro2}
\left(\begin{array}{cc}
  \renewcommand{\arraystretch}{0.6}
  \scriptscriptstyle
    \left(\ve_{\beta\beta'}(n)
    \frac{\tau_{m,n+e_\beta-e_{\beta'}}
    (t_{\beta'}-[z^{-1}])} {\tau_{mn}}z^{n_{\beta'}+\delta_{\beta\beta'}-1}
     \right)_{\begin{array}{c}
      \scriptscriptstyle 1\leq \beta\leq p\\
      \scriptscriptstyle 1\leq \beta'\leq p
     \end{array}}
  &
  \renewcommand{\arraystretch}{0.6}
  \scriptstyle
  \left(\e_{\a\beta}(m,n)
  \frac{\tau_{m+e_\a,n+e_\beta}(s_{\a}-[z^{-1}])}
  {\tau_{mn}}z^{-m_{\a}-1}
     \right)_{\begin{array}{c}
      \scriptscriptstyle 1\leq \beta\leq p\\
      \scriptscriptstyle 1\leq \a\leq q
     \end{array}}\\
  \renewcommand{\arraystretch}{0.6}
  \scriptstyle
  \left(\e_{\a\beta}(m,n)\frac{\tau_{m-e_\a,n-e_\beta}
  (t_{\beta}-[z^{-1}])} {\tau_{mn}}z^{n_{\beta}-1}\right)
  _{\begin{array}{c}
      \scriptscriptstyle 1\leq \a\leq q\\
      \scriptscriptstyle 1\leq \beta\leq p
     \end{array}}
  &
  \renewcommand{\arraystretch}{0.6}
  \scriptstyle
  \left(\ve_{\a'\a}(m)
  \frac{\tau_{m+e_\a-e_{\a'},n}(s_{\a}-[z^{-1}])}
   {\tau_{mn}}z^{\delta_{\a\a'}-1-m_{\a}}
     \right)_{\begin{array}{c}
      \scriptscriptstyle 1\leq \a'\leq q\\
      \scriptscriptstyle 1\leq \a\leq q
     \end{array}}
\end{array}\right),
\end{equation}%
with inverse transpose matrix given by
\begin{equation*}
  W_{mn}^*(z)\diag \left(e^{\sum_1^{\iy}t_{1k}z^k},\ldots,e^{\sum_1^{\iy}t_{pk}z^k},
  e^{\sum_1^{\iy}s_{1k}z^k},\ldots,e^{\sum_1^{\iy}s_{qk}z^k}\right)=
 \hspace*{3cm} \renewcommand{\arraystretch}{2}
\end{equation*}%
\begin{equation}\label{intro3}
\left(\begin{array}{cc}
  \renewcommand{\arraystretch}{0.6}
  \scriptscriptstyle
    \left(\ve_{\beta'\beta}(n)
    \frac{\tau_{m,n+e_\beta-e_{\beta'}}(t_{\beta}+[z^{-1}])}
     {\tau_{mn}}z^{\delta_{\beta'\beta}-1-n_{\beta}}
     \right)_{\begin{array}{c}
      \scriptscriptstyle 1\leq \beta'\leq p\\
      \scriptscriptstyle 1\leq \beta\leq p
     \end{array}}
  &
  \renewcommand{\arraystretch}{0.6}
  \scriptstyle
  \left(-\e_{\beta\a}(n,m)\frac{\tau_{m-e_\a,n-e_\beta}
  (s_{\a}+[z^{-1}])} {\tau_{mn}}z^{m_{\a}-1}
     \right)_{\begin{array}{c}
      \scriptscriptstyle 1\leq \beta\leq p\\
      \scriptscriptstyle 1\leq \a\leq q
     \end{array}}\\
  \renewcommand{\arraystretch}{0.6}
  \scriptstyle
  \left(-\e_{\beta\a}(n,m)\frac{\tau_{m+e_\a,n+e_\beta}
  (t_{\beta}+[z^{-1}])} {\tau_{mn}}z^{-n_{\beta}-1}\right)
  _{\begin{array}{c}
      \scriptscriptstyle 1\leq \a\leq q\\
      \scriptscriptstyle 1\leq \beta\leq p
     \end{array}}
  &
  \renewcommand{\arraystretch}{0.6}
  \scriptstyle
  \left(\ve_{\a\a'}(m)
  \frac{\tau_{m+e_\a-e_\a',n}(s_{\a'}+[z^{-1}])}
   {\tau_{mn}}z^{\delta_{\a\a'}-1+m_{\a'}}
     \right)_{\begin{array}{c}
      \scriptscriptstyle 1\leq \a\leq q\\
      \scriptscriptstyle 1\leq \a'\leq q
     \end{array}}
\end{array}\right).
\end{equation}%
The matrices $W_{mn}(z)$ and $W_{m^*n^*}^*(z)$ satisfy the bilinear identities {\em which characterize the
$\tau$-function of the $(p+q)$-KP hierarchy}
\begin{equation}\label{intro4}
  \oint_\infty W_{mn}(z;s,t)W^*_{m^*n^*}(z;s^*,t^*)^\top dz=0,
\end{equation}
for all $m,n,m^*,n^*$ such that $\vert m\vert=\vert n\vert$, $\vert m^*\vert=\vert n^*\vert$ and all $s,t,s^*,t^*
\in \C^{\iy}$. The integral above is taken along a small circle about $z=\iy$; writing out the identity above
componentwise and using the expressions (\ref{intro2}) and (\ref{intro3}) for $W$ and $W^*$, the bilinear identity
(\ref{intro4}) is equivalent to the single identity
\begin{eqnarray*}
  \sum_{\beta=1}^p\oint_\infty
  \scriptstyle(-1)^{\sigma_\beta(n)}\,\tau_{m,n-e_{\beta}}
    (t_{\beta}-[z^{-1}])\tau_{m^*,n^*+e_{\beta}}(t^*_{\beta}+[z^{-1}])
          e^{\scriptscriptstyle\sum_1^{\iy} (t_{\beta k}-t_{\beta k}^*)z^k}z^{n_\beta-n_\beta^*-2}\,dz=\nonumber\\
  \sum_{\a=1}^q\oint_\infty  \scriptstyle (-1)^{\sigma_\a(m)}\,\tau_{m+e_\a,n}(s_{\a}-[z^{-1}])\tau_{m^*-e_\a
    ,n^*}(s^*_{\a}+[z^{-1}])e^{\sum_1^{\iy} (s_{\a k}-s_{\a k}^*)z^k}\,z^{m_{\a}^*-m_\a-2}\,dz,
\end{eqnarray*}
  where $\vert m^*\vert=\vert n^*\vert+1$ and $\vert m\vert=\vert
  n\vert-1$ and
   \begin{equation*}
     \sigma_\a(m)={\sum_{\a'=1}^\a(m_{\a'}
-m_{\a'}^*)}\quad\hbox{and}\quad
     \sigma_\beta(n)={\sum_{\beta'=1}^\beta(n_{\beta'}-n_{\beta'}^*)}.
   \end{equation*}%
   It remains an open problem to have a clear
   understanding of why the $W_{mn}(z;s,t)$-matrix above,
   evaluated at $t=s=0$, coincides with the
   Riemann-Hilbert matrix for the
   mixed multiple orthogonal polynomials.
\bigskip

\noindent{\bf PDE's for the determinant of moment matrices}: Upon actually computing the residues in the contour
integrals above, the functions $\tau_{mn}$, with $\vert m\vert=\vert n\vert$, satisfy the following PDE's expressed
in terms of the Hirota symbol\footnote{For a given polynomial $p(t_1,t_2,\dots)$, the Hirota symbol between functions
$f=f(t_1,t_2,\ldots)$ and $g=g(t_1,t_2,\ldots)$ is defined by:
$$
  p(\frac{\p}{\p t_1},\frac{\p}{\p t_2},\dots)f\circ g:= p(\frac{\p}{\p y_1},\frac{\p}{\p y_2},\dots)f(t+y)g(t-y)\Bigl|_{y=0}.
$$
We also need the elementary Schur polynomials ${S}_{\ell}$, defined by $e^{\sum^{\iy}_{1}t_kz^k}:=\sum_{k\geq 0} {
S}_k(t)z^k$ for $\ell\geq 0$ and ${S}_{\ell}(t)=0$ for $\ell<0$; moreover, set
$$
  S_{\ell}(\tilde \p_t):={S}_{\ell}(\frac{\p}{\p t_1},\frac{1}{2}\frac{\p}{\p t_2},\frac{1}{3}\frac{\p}{\p t_3},\ldots).
$$
}:
\begin{eqnarray}\label{introa}
  \tau_{mn}^2\frac{\p^2}{\p t_{\beta,\ell+1}\p t_{\beta',1}}\ln \tau_{mn}
  &=& S_{\ell+2\delta_{\beta\beta'}}\bigl(\tilde\p_{t_{\beta}}\bigr)
     \tau_{m,n+e_{\beta}-e_{\beta'}}\circ\tau_{m,n+e_{\beta'}-e_{\beta}}\nonumber\\
  \tau_{mn}^2 \frac{\p^2}{\p s_{\a,\ell+1}\p s_{\a',1}}\ln \tau_{mn}
  &=&S_{\ell+2\delta_{\a\a'}}(\tilde\partial_{s_{\a}})
     \tau_{m+e_{\a'}-e_{\a},n}\circ\tau_{m+e_{\a}-e_{\a'},n}\nonumber\\
  \tau_{mn}^2 \frac{\p^2}{\p s_{\a,1}\p t_{\beta,\ell+1}}\ln \tau_{mn}
  &=&-S_{\ell}(\tilde\partial_{t_{\beta}})\tau_{m+e_{\a},n+e_\beta}\circ\tau_{m-e_{\a},n-e_{\beta}}\nonumber\\
  \tau_{mn}^2\frac{\p^2}{\p t_{\beta,1}\p s_{\a,\ell+1}}\ln \tau_{mn}
  &=&-S_{\ell}(\tilde\partial_{s_{\a}})\tau_{m-e_{\a},n-e_\beta}\circ\tau_{m+e_{\a},n+e_{\beta}}.
\end{eqnarray}
Whereas the formulae above have in their right hand side different $\tau_{mn}$'s, one can combine these relations
to yield PDE's in a single $ \tau_{mn}$; so, these are PDE's for the determinant of the moment matrix
(\ref{intro1}). In particular, one finds the following ${p+q}\choose 2$ PDE's, which play a fundamental role in
chains of random matrices and in the transition probabilities for critical infinite-dimensional diffusions:
\begin{eqnarray*}
  \pp{}{t_{\beta',1}}\left(\frac{\frac{\p^2}{\p t_{\beta,2}\p t_{\beta',1}}\ln\tau_{mn}}{\frac{\p^2}{\p t_{\beta,1}\p
      t_{\beta',1}}\ln\tau_{mn}}\right)+
  \pp{}{t_{\beta,1}}\left(\frac{\frac{\p^2}{\p t_{\beta',2}\p t_{\beta,1}}\ln\tau_{mn}}{\frac{\p^2}{\p t_{\beta',1}\p
      t_{\beta,1}}\ln\tau_{mn}}\right)&=&0,\\
  \pp{}{s_{\a',1}}\left(\frac{\frac{\p^2}{\p s_{\a,2}\p s_{\a',1}}\ln\tau_{mn}}{\frac{\p^2}{\p s_{\a,1}\p
      s_{\a',1}}\ln\tau_{mn}}\right)+
  \pp{}{s_{\a,1}}\left(\frac{\frac{\p^2}{\p s_{\a',2}\p s_{\a,1}}\ln\tau_{mn}}{\frac{\p^2}{\p s_{\a',1}\p
      s_{\a,1}}\ln\tau_{mn}}\right)&=&0,\\
  \pp{}{s_{\a,1}}\left(\frac{\frac{\p^2}{\p t_{\beta,2}\p s_{\a,1}}\ln\tau_{mn}}{\frac{\p^2}{\p t_{\beta,1}\p
      s_{\a,1}}\ln\tau_{mn}}\right)  +
  \pp{}{t_{\beta,1}}\left(\frac{\frac{\p^2}{\p s_{\a,2}\p t_{\beta,1}}\ln\tau_{mn}}{\frac{\p^2}{\p s_{\a,1}\p
      t_{\beta,1}}\ln\tau_{mn}}\right)&=&0.
\end{eqnarray*}

\section{Tau functions and mixed multiple orthogonal polynomials}
Following \cite{DK} we introduce the notion of mixed
multiple orthogonal polynomials (mixed mops), with
regard to two sets of weights
$\set{\rhi_1,\rhi_2,\dots,\rhi_p}$ and
$\set{\psi_1,\psi_2,\dots,\psi_q}$:
%
\begin{definition}
  Let $A_1,\,A_2,\dots,A_p$ be $p$ polynomials in $y$ and set%
  \begin{equation*}
    Q(y):=A_1(y)\rhi_1(y)+A_2(y)\rhi_2(y)+\cdots +A_p(y)\rhi_p(y).
  \end{equation*}%
  {\bf Type I}\quad For $\a\in\set{1,2,\dots,q}$ the polynomials $A_1,\,A_2,\dots,A_p$ are said to be \emph{Type I
  normalized} with respect to $\psi_\a$, denoted \Type {I}{\psi_\a}, if $\deg (A_\b)<n_\b$ for $\b=1,\dots,p$ and
  $Q$ satisfies the following orthogonality conditions
  \begin{equation}\label{mmops_cond_i}
    \inn{x^i\psi_{\a'}(x)}{Q(y)}=\delta_{\a\a'}\delta_{i,m_\a-1},\qquad i=0,\dots,m_{\a'}-1,\ 1\leq\a'\leq q.
  \end{equation}
  {\bf Type II}\quad For $\b\in\set{1,\dots,p}$ the polynomials $A_1,\,A_2,\dots,A_p$ are said to be \emph{Type II
  normalized} with respect to $\rhi_\b$, denoted \Type {II}{\rhi_\b}, if $A_\b$ is monic of degree $n_\b$ and
  $\deg(A_{\b'})<n_{\b'}$ for $1\leq\b'\leq p$, with $\b'\neq\b$, and $Q$ satisfies the following orthogonality
  conditions
  \begin{equation}\label{mmops_cond_ii}
    \inn{x^i\psi_\a(x)}{Q(y)}=0,\qquad i=0,\dots,m_\a-1,\ 1\leq \a\leq q.
  \end{equation}
  \par\smallskip\noindent
  In both cases, the polynomials $A_1,\dots,A_p$ are called \emph{multiple orthogonal polynomials of mixed type},
  or \emph{mixed mops} for brevity.
\end{definition}
\begin{proposition}\label{prp:typeII}
   For $\b=1,\dots,p$, let
  \begin{equation}\label{Q1Q2}
    \Q{\b}(y):=\Q{\b,1}(y)\rhi^t_1(y)+\cdots+\Q{\b,p}(y)\rhi^t_p(y),
  \end{equation}%
  where $\Q{\b,\b'}$, with $1\leq \b,\b'\leq p$  are the polynomials, defined by
  \begin{equation}
    \begin{array}{rcl}
      \Q{\b,\b}(z)&:=&\ds z^{n_\b}\frac{\tau_{mn}(t_\b-\zs)}{\tau_{mn}}\\
      \Q{\b,\b'}(z)&:=&\ds\ve_{\b\b'}(n)z^{n_{\b'}-1}\frac{\tau_{m,n+e_\b-e_{\b'}}(t_{\b'}-\zs)}{\tau_{mn}},\qquad \b'\neq\b,
    \end{array}
  \end{equation}
  and
  \begin{equation}\label{eps_def}
    \renewcommand{\arraystretch}{1.3}
    \ve_{\b\b'}(n)=\left\{
    \begin{array}{lll}
      (-1)^{n_{\b'+1}+n_{\b'+2}+\cdots+n_{\b}+1}&\hbox{ if }&\b>\b',\\
      (-1)^{n_{\b+1}+n_{\b+2}+\cdots+n_{\b'}}&\hbox{ if }&\b<\b'.
    \end{array}
    \right.
  \end{equation}%
  Then $\Q{\b,1}(y),\dots,\Q{\b,p}(y)$ are \Type {II}{\rhi^t_\b} mixed mops.
\end{proposition}
\goodbreak
\begin{proof}\quad
For $j=0,1,2,\dots$ and $\b=1,2,\dots,p$ we define a column vector $C_j^\b$ of size $\vert m\vert$ by
\begin{equation}\label{eq:Cjb}
  C_j^\b:=
  \renewcommand{\arraystretch}{1.8}
  \begin{matrix}{c}
    \Bigl(\inn{x^{i_1}\psi^{-s}_1(x)}{y^{j}\rhi^t_\b(y)}\Bigr)_{0\leq i_1<m_1}\\
    \Bigl(\inn{x^{i_2}\psi^{-s}_2(x)}{y^{j}\rhi^t_\b(y)}\Bigr)_{0\leq i_2<m_2}\\
    \vdots\\
    \Bigl(\inn{x^{i_q}\psi^{-s}_q(x)}{y^{j}\rhi^t_\b(y)}\Bigr)_{0\leq i_q<m_q}
  \end{matrix}.
\end{equation}%
When its size is important (see the proof of Proposition
\ref{prp:typeI}) we write $C_j^\b(m)$ for
(\ref{eq:Cjb}). Notice that the moment matrix $T_{mn}$,
defined in (\ref{Tmn_def})
,
can be expressed in terms of the columns $C_j^\b$, and
so
\begin{equation}\label{tau_C}
  \tau_{mn}=\det\left(C_0^1,\ C_1^1,\ \dots,\ C_{n_1-1}^1,\ C_0^2,\ C_1^2,\ \dots,\ C_{n_p-1}^{p}\right).
\end{equation}%
For future use, let us point out that the dependence of $\tau_{mn}$ on the $t$ variables is as follows:
\begin{eqn}{tau_Ct}
  \tau_{mn}(t_1)&=&\det(C_0^1(t_1),\ C_1^1(t_1),\ \dots,\ C^1_{n_1-1}(t_1),\ C_0^2,\ C_1^2,\ \dots,\ C_{n_p-1}^{p}),\\
  \tau_{mn}(t_2)&=&\det(C_0^1,\ C_1^1,\ \dots,\ C^1_{n_1-1},\ C_0^2(t_2),\ C_1^2(t_2),\ \dots,\ C_{n_p-1}^{p}),\\
  &\vdots&\\
  \tau_{mn}(t_{n_p})&=&\det(C_0^1,\ C_1^1,\ \dots,\ C^1_{n_1-1},\ C_0^2,\ C_1^2,\ \dots,\ C_{n_p-1}^{p}(t_{n_p})).
\end{eqn}%
Since
\begin{equation}\label{Q_1_def}
  \Q{\b}(y)=\ds\sum_{\b'=1}^p \Q{\b,\b'}(y)\rhi^t_{\b'}(y)=
  \ds\sum_{\b'=1}^p \left(\delta_{\b\b'}y^{n_\b}+\sum_{j=0}^{n_{\b'}-1}A_{\b\b'}^jy^j\right)\rhi^t_{\b'}(y),
\end{equation}%
the orthogonality conditions (\ref{mmops_cond_ii}) for $Q=\Q{\b}$ can be written as the linear system
\begin{equation*}
  \sum_{\b'=1}^p\sum_{j=0}^{n_{\b'-1}}A_{\b\b'}^jC_j^{\b'}=-C_{n_\b}^\b,
\end{equation*}%
of $\vert m\vert$ equations, in the $\vert n\vert$ ($=\vert m\vert$) unknowns $A_{\b\b'}^j$, where $1\leq \b'\leq
p$ and $0\leq j<n_{\b'-1}$. If we order these unknowns as follows:
$A_{\b1}^0,\,A_{\b1}^1,\dots,\A_{\b1}^{n_1-1},\,A_{\b2}^0,$ $A_{\b2}^{1},\dots,A_{\b p}^{n_p-1}$, then this linear
system has precisely $\tau_{mn}$ as determinant, in view of (\ref{tau_C}). Since $\tau_{mn}\neq0$, generically, we
have by Cramer's rule,
\begin{equation}\label{eq:Abb'}
  A_{\b\b'}^j=\frac{\det\left(C_0^1,\, C_1^1,\, \dots,C_{j-1}^{\b'},\, -C_{n_\b}^\b,\, C_{j+1}^{\b'},\, \dots,\
          C^p_{n_p-1}\right)}{\tau_{mn}}.
\end{equation}%
Substituted in (\ref{Q_1_def}) this yields an explicit expression for the \Type {II}{\rhi^t_\b} mixed mops
$\Q{\b,1}(y),\dots,\Q{\b,p}(y)$.

\smallskip

In order to connect these polynomials with the tau functions $\tau_{mn}$ we first expand $\tau_{mn}(t_\b-\zs)$
using (\ref{tau_Ct}). Thus, we need to compute $C_j^\b(t_\b-\zs)$, which we claim to be given by
\begin{equation}\label{C_shift}
  C_j^\b(t_\b-\zs)=C_j^\b(t_\b)-z^\mi C^\b_{j+1}(t_\b)
  =C_j^\b-z^\mi C^\b_{j+1},
\end{equation}%
where the last equality is the notational simplification
agreed upon. To prove the first equality in
(\ref{C_shift}), which is an equality of formal series
in $z^{-1}$, let us write a typical entry of the column
vector $C_j^\b(t_\b)$ with its explicit time-dependence
on $t_\b$,
\begin{equation*}
  \inn{x^{i}\psi^{-s}_\a(x)}{y^j\rhi^t_\b(y)}=
   \inn{x^{i}\psi^{-s}_\a(x)}{y^j\rhi_\b(y)
   e^{\sum_{k=1}^\iy t_{\beta k}y^k}}
   ,
\end{equation*}%
where $1\leq\a\leq q$ and $0\leq i<m_\a$.
 The following trivial identity will be used over and over
 again in this paper
 \begin{equation}
 e^{-\sum_1^\iy \frac{x^i}{i}}=1-x.
 \label{transl}\end{equation}
 In view of
the latter, the same entry of $C_j^\b(t_\b-\zs)$ (as
above) is given by
%
\begin{equation*}
  \inn{x^i\psi^{-s}_\a(x)}{y^j\rhi^t_\b(y)\left(1-\frac yz\right)}=\inn{x^i\psi^{-s}_\a(x)}{y^j\rhi^t_\b(y)}-
         \frac1z\inn{x^i\psi^{-s}_\a(x)}{y^{j+1}\rhi^t_\b(y)}
\end{equation*}%
which proves (\ref{C_shift}). Using the fact that the determinant is a skew-symmetric multilinear function of its
columns, which vanishes when two columns are equal, it follows from (\ref{tau_Ct}), (\ref{C_shift}) and
(\ref{eq:Abb'}) that
\begin{eqnarray*}
  \lefteqn{z^{n_\b}\tau_{mn}(t_\b-\zs)}\\
  &=&\det(C_0^1,\dots,C_{n_{\b-1}-1}^{\b-1},\,zC_0^\b-C_1^\b,\dots,zC_{n_\b-1}^\b-C_{n_\b}^\b,\,C_0^{\b+1},\dots,C^p_{n_p-1})\\
  &\stackrel{(*)}=&\sum_{j=0}^{n_\b}z^j\det(C_0^1,\,C_1^1,\dots,C_{j-1}^\b,-C_{j+1}^\b,\dots,-C_{n_\b}^\b,\,
                    C_0^{\b+1},\dots,C^p_{n_p-1})\\
  &=&\sum_{j=0}^{n_\b}z^j\det(C_0^1,\,C_1^1,\dots,C_{j-1}^\b,\,-C_{n_\b}^\b,\,C_{j+1}^\b,\dots,C_{n_\b-1}^\b,\,
              C_0^{\b+1},\dots,C^p_{n_p-1})\\
  &=&\sum_{j=0}^{n_\b}z^jA_{\b\b}^j\tau_{mn}\\
  &=&\tau_{mn}\,\Q{\b,\b}(z).
\end{eqnarray*}
In $(*)$ it is understood that all the columns between
$-C_{j+1}^\b$ and $-C_{n_\b}^\b$ come with negative signs
and no others; this notation shall be used freely in the sequel, without further mention.

\smallskip

For $\Q{\b,\b'}$ with $\b\neq\b'$ we also need to keep track of signs and of shifts in the first index of the tau
function, as is seen in the following computation, where we suppose that $\b<\b'$:
\begin{eqnarray*}
  \lefteqn{z^{n_{\b'-1}}\tau_{m,n+e_\b-e_{\b'}}(t_{\b'}-\zs)}\\
  &=&\det(C_0^1,\dots,C_{n_\b-1}^{\b},\,C_{n_\b}^\b,\dots,zC^{\b'}_0-C^{\b'}_1,\dots,zC^{\b'}_{n_{\b'}-2}-C^{\b'}_{n_{\b'}-1},
        C_0^{\b'+1},\dots,C^p_{n_p-1})\\
  &=&\sum_{j=0}^{n_{\b'}-1}z^{j}\det(C_0^1,\dots,\,C_{n_\b}^\b,\dots,C_{j-1}^{\b'},\,-C^{\b'}_{j+1},\dots,
        -C^{\b'}_{n_{\b'-1}},\,C_0^{\b'+1},\dots,C_{n_p-1}^p)\\
  &=&\ve_{\b\b'}(n)\sum_{j=0}^{n_{\b'}-1}z^{j}\det(C_0^1,\dots,\,C_{n_{\b-1}}^\b,C_0^{\b+1},\dots,C_{j-1}^{\b'},\,-C_{n_\b}^\b,\,
         C^{\b'}_{j+1},\dots,C_{n_p-1}^p)\\
  &=&\ve_{\b\b'}(n)\sum_{j=0}^{n_{\b'}-1}z^{j}\,A_{\b\b'}^j\,\tau_{mn}\\
  &=&\ve_{\b\b'}(n)\,\tau_{mn}\,\Q{\b,\b'}(z).
\end{eqnarray*}
The sign $\ve_{\b\b'}(n)$ which we introduced when
moving the column $C_{n_\b}^\b$ to the right is given by
$(-1)^{n_{\b+1}+\cdots+n_{\b'}}$, in agreement with
(\ref{eps_def}). When $\b>\b'$ the column $C_{n_\b}^\b$
is moved to the left, which yields a sign
$\ve_{\b\b'}(n)=-(-1)^{n_{\b'+1}+\cdots+n_{\b}}$, as is
easily checked.
\qed
\end{proof}

The tau functions $\tau_{mn}$ also lead to Type I normalized mixed mops, as given in the following proposition.
\begin{proposition}\label{prp:typeI}
   For $\a=1,\dots,q,$ let
  \begin{equation}\label{R1R2}
    \P{\a}(y):=\P{\a,1}(y)\rhi^t_1(y)+\cdots+\P{\a,p}(y)\rhi^t_{p}(y),
  \end{equation}%
  where $\P{\a,\b}$ are the polynomials, defined by
  \begin{equation}
    \P{\a,\b}(z):=\e_{\a\b}(m,n)z^{n_\b-1}\frac{\tau_{m-e_\a,n-e_\b}(t_\b-\zs)}{\tau_{mn}},
  \end{equation}
  with leading sign
  \begin{equation}\label{eps2_def}
    \e_{\a\b}(m,n)=(-1)^{m_1+\cdots+m_\a}(-1)^{n_1+\cdots+n_\b}.
  \end{equation}
  Then $\P{\a,1}(y),\dots,\P{\a,p}(y)$ are \Type {I}{\psi^{-s}_\a} mixed mops.
\end{proposition}
\begin{proof}
\quad Letting
\begin{equation}\label{R_1_def}
  \P{\a}(y)=\sum_{\b=1}^p\P{\a,\b}(y)\rhi^t_\b(y)=\sum_{\b=1}^p\sum_{j=0}^{n_\b-1}B_{\a\b}^j\,y^j\,\rhi^t_\b(y),
\end{equation}%
the orthogonality conditions (\ref{mmops_cond_i}) for $Q=\P{\a}$ can be written as the linear system
\begin{equation*}
  \sum_{\b=1}^p\sum_{j=0}^{m_\b-1}B_{\a\b}^jC_j^\b=E_{m_\a}^\a,
\end{equation*}%
where $E_{m_\a}^\a$ denotes the column vector of size $\vert m\vert$ with a $1$ at position $m_\a$ of the $\a$-th
block (so at position $m_1+m_2+\cdots+m_\a$), and zeros elsewhere. Cramer's rule now yields
\begin{eqn*}[2.5]
  B^j_{\a\b}&=&\ds\frac{\det(C_0^1,\, C_1^1,\dots,C^\b_{j-1},\, E^\a_{m_\a},\, C_{j+1}^\b,\dots,C^p_{n_p-1})}{\tau_{mn}}, \\
     &=&\ds\e_{\a\b}(m,n)(-1)^{j+1-n_\b}\frac{\det(D_0^1,\, D_1^1,\dots,D^\b_{j-1},\, \widehat{D^\b_j},\,
              D_{j+1}^\b,\dots,D^p_{n_p-1})} {\tau_{mn}},
\end{eqn*}%
where the last line was obtained by expanding the
determinants along the $E_{m_\a}^\a$ column,
$\e_{\a\b}(m,n)$ is given by (\ref{eps2_def}) and
$D_k^\gamma$ is the column vector $D_k^\gamma$ with its
$(m_1+\cdots+m_\a)$-th entry removed, i.e.,
$D_k^\gamma:=C_k^\gamma(m-e_\a)$. This yields explicit
expressions for the \Type {I}{\psi^{-s}_\a} mixed mops.
To connect them with tau functions, we notice on the one
hand that the columns $D_k^\gamma$ appear in the
matrices which define the tau functions
$\tau_{m-e_\a,\star}$, and on the other hand that these
columns behave in the same way (\ref{C_shift}) as
$C_k^\gamma$ under shifts. Therefore we can compute, as
before
\begin{eqnarray*}
  \lefteqn{z^{n_\b-1}\tau_{m-e_\a,n-e_\b}(t_\b-\zs)}\\
  &=&\det(D_0^1,\dots,D_{n_{\b-1}-1}^{\b-1},\,zD_0^\b-D_1^\b,\dots,zD_{n_\b-2}^\b-D_{n_\b-1}^\b,D_0^{\b+1},\dots,D_{n_p-1}^p)\\
  &=&\sum_{j=0}^{n_\b-1}z^j\det(D_0^1,\dots,D_{j-1}^\b,\,-D_{j+1}^\b,\dots,-D_{n_\b-1}^\b,\,D_0^{\b+1},\dots,D_{n_p-1}^p)\\
  &=&\sum_{j=0}^{n_\b-1}(-1)^{n_\b-j-1}z^j\det(D_0^1,\dots,D_{j-1}^\b,\,\widehat{D_j^\b},\,D_{j+1}^\b,\dots,D_{n_p-1}^p)\\
  &=&\e_{\a\b}(m,n)\sum_{j=0}^{n_\b-1}z^j\,B_{\a\b}^j\,\tau_{mn}\\
  &=&\e_{\a\b}(m,n)\,\P{\a,\b}(z)\,\tau_{mn}.
\end{eqnarray*}
\qed
\end{proof}

\section{Cauchy transforms}\label{par:cauchy}
We now show that certain shifts of the tau function,
appearing in the Riemann-Hilbert matrix of \cite{DK},
are (formal) Cauchy transforms. For a function $F$ and a
weight $\psi$, define its Cauchy transform as
\begin{equation}\label{eq:cauchy_def}
  \CC_\psi G(z):=
  \inn{\frac{\psi(x)}{z-x}}{G(y)}=\sum_{i=0}^\infty\frac1{z^{i+1}}\inn{x^i\psi(x)}{G(y)},
\end{equation}%
i.e., our Cauchy transforms will be formal in the sense
that we always think of $z$ as being large, and this is
precisely how it will be used. The first type of Cauchy
transforms which we are interested in are given in the
following proposition.
\begin{proposition}\label{prp:cauchy_q}
  For $\a=1,\dots,q$ and $\b=1,\dots,p$, the Cauchy transforms of
  $\Q{\b}(y)=\Q{\b,1}(y)\rhi^t_1(y)+\cdots+\Q{\b,p}(y)\rhi^t_p(y),$ with respect to $\psi^{-s}_\a$ can be expressed in terms
  of tau functions as follows.
  \begin{equation}\label{for:cauchy}
    \CC_{\psi^{-s}_\a}\Q{\b}(z)=\e_{\a\b}(m,n)\,z^{-m_\a-1}\,\frac{\tau_{m+e_\a,n+e_\b}(s_\a-\zs)}{\tau_{mn}}.
  \end{equation}%
\end{proposition}
\begin{proof}\quad
The proof is based on an investigation of the moment matrix by row. Therefore we define, for $\a=1,\dots,q$ and for
$i=0,1,2,\dots$ the row $R_\a^i$ of size~$\vert n\vert$ by
\begin{equation*}
  R_\a^i:=
  \begin{matrix}{c}
    \left(\inn{x^{i}\psi^{-s}_\a(x)}{y^{j_1}\rhi^t_1(y)}\right)_{0\leq j_1<n_1}\ \cdots\
    \left(\inn{x^{i}\psi^{-s}_\a(x)}{y^{j_p}\rhi^t_p(y)}\right)_{0\leq j_p<n_p}
  \end{matrix}.
\end{equation*}%
When its size is important we write $R_\a^i(n)$ for $R_\a^i(n)$. The moment matrix $T_{mn}$ can now be
expressed in terms of the rows $R_\a^i$, and so
\begin{equation}\label{tau_R}
  \renewcommand{\arraystretch}{1.2}
  \tau_{mn}=\det
  \begin{matrix}{c}
    R_1^0\\R_1^1\\ \vdots\\R_1^{m_1-1}\\ R_2^0\\ \vdots\\ R_q^{m_q-1}
  \end{matrix}.
\end{equation}
The tau function which we need to compute is
$\tau_{m+e_\a,n+e_\b}(s_\a-\zs)$; so throughout the
proof, $R_{\a'}^i$ stands for $R_{\a'}^i(n+e_\b)$ for
all $1\leq \a'\leq q$ and $i=0,1,2,\dots$. Notice that
the only rows which depend on the time variables
$s_\a=(s_{\a1},\,s_{\a2},\dots)$ are the rows $R_\a^i$.
Recall the dependence of $\psi^{-s}_\a$ on $s_\a$ as
follows
\begin{equation*}
  \psi_\a^{-s}(x)=\psi_\a(x)e^{-\sum_{k=1}^\iy s_{\a k}x^k}%
  ,
\end{equation*}%
so that, according to the identity (\ref{transl}), when
$s_\a$ gets replaced by $s_\a-\zs$, then
$\psi^{-s}_\a(x)$ gets replaced by
$\psi^{-s}_\a(x)\left(1+\frac
xz+\frac{x^2}{z^2}+\cdots\right).$ It follows that
\begin{equation*}
  \renewcommand{\arraystretch}{0.6}
  R_\a^i(s_\a-\zs)=\left(\inn{x^{i}\psi^{-s}_\a(x)
           \left(1+\frac xz+\frac{x^2}{z^2}+\cdots\right)}{
           \begin{array}{l} \\ \\ \end{array}\!\! y^{j_{\b'}}\rhi^t_{\b'}(y)}\right)_{
    \begin{array}{l}
      \scriptstyle 1\leq \b'<p\\
      \scriptstyle 0\leq j_{\b'}<n'_{\b'}
     \end{array}},
\end{equation*}%
where we introduce the convenient abbreviation
$n_{\b'}':=n_{\b'}+\delta_{\b\b'}=(n+e_\b)_{\b'}$.
Notice that
\begin{equation*}
  R_\a^i(s_\a-\zs)=R_\a^i(s_\a)+\frac1zR_\a^{i+1}(s_\a-\zs),
\end{equation*}%
for $0\leq i\leq m_\a-1$; we stop at $m_\a-1$ because the highest index $i$ for which $R_\a^i$ appears in
$T_{m+e_\a,n+e_\b}$ is $i=m_\a$. By recursively applying this formula we get that for $0\leq i\leq m_\a-1$
\begin{equation*}
  R_\a^i(s_\a-\zs)=R_\a^i(s_\a)+\hbox{ lin.\ comb.\ of lower rows}.
\end{equation*}%
This leads to the first equality in
\begin{eqnarray}\label{for:first_eq}
  \lefteqn{z^{-m_\a-1}\tau_{m+e_\a,n+e_\b}(s_\a-\zs)}\hskip1cm\\
  &=&z^{-m_\a-1}\det
  \renewcommand{\arraystretch}{1.2}
  \begin{matrix}{c}
    R_1^0\\ \vdots\\ R_\a^{m_\a-1}\\ R_\a^{m_\a}(s_\a-\zs)\\ R_{\a+1}^0\\ \vdots\\ R_q^{m_q-1}
  \end{matrix}=\det
  \begin{matrix}{c}
    R_1^0\\ \vdots\\ R_\a^{m_\a-1}\\ \tilde R_\a(z)\\ R_{\a+1}^0\\ \vdots\\ R_q^{m_q-1}
  \end{matrix}.\nonumber
\end{eqnarray}%
For the second equality, in which we have put
\begin{equation}\label{eq:Rtilde}
  \tilde R_\a(z):=
  \begin{matrix}{c}
    \left(\inn{\frac{\psi^{-s}_\a(x)}{z-x}}{y^{j_1}\rhi^t_1(y)}\right)_{0\leq j_1<n_1'}\ \cdots\
    \left(\inn{\frac{\psi^{-s}_\a(x)}{z-x}}{y^{j_p}\rhi^t_p(y)}\right)_{0\leq j_p<n_p'}
  \end{matrix},
\end{equation}%
it suffices to show that
\begin{equation}\label{for:higher}
  R_\a^{m_\a}(s_\a-\zs)=z^{m_\a+1}\tilde R_\a(z)+\hbox{ lin.\ comb.\ of higher rows $R_\a^i$.}
\end{equation}%
To do this, compare a typical entry of $R_\a^{m_\a}(s_\a-\zs)$, to wit
\begin{eqnarray*}
  \lefteqn{\inn{x^{m_\a}\psi^{-s}_\a(x)\left(1+\frac xz+\frac{x^2}{z^2}+\cdots\right)}{y^{j}\rhi^t_{\b'}(y)}}\\
      &=&z^{m_\a}\inn{\left(\frac xz\right)^{m_\a}\psi^{-s}_\a(x)\left(1+\frac xz+\frac{x^2}{z^2}+
             \cdots\right)}{y^{j}\rhi^t_{\b'}(y)}\\
      &=&z^{m_\a}\inn{\psi^{-s}_\a(x)\left(\left(\frac xz\right)^{m_\a}+\left(\frac
                    xz\right)^{m_\a+1}+\cdots\right)}{y^{j}\rhi^t_{\b'}(y)}
\end{eqnarray*}
with the corresponding typical entry of $z^{m_\a+1}\tilde R_\a(z)$, to wit
\begin{eqnarray*}
  \lefteqn{z^{m_\a+1}\inn{\frac{\psi^{-s}_\a(x)}{z-x}}{y^{j}\rhi^t_{\b'}(y)}}\\
  &=&z^{m_\a}\inn{\psi^{-s}_\a(x)\left(1+\frac xz+\left(\frac xz\right)^2+\cdots\right)}{y^{j}\rhi^t_{\b'}(y)}.
\end{eqnarray*}
It leads to the following explicit expression for (\ref{for:higher}):
\begin{equation*}
  R_\a^{m_\a}(s_\a-\zs)+\sum_{i=0}^{m_\a-1}z^{m_\a-i}R_\a^i(s_\a)=z^{m_\a+1}\tilde R_\a(z),
\end{equation*}%
and hence to the proof of the second equality in (\ref{for:first_eq}).

\smallskip

In order to make the connection with mixed mops, we introduce for $\b'=1,\dots,p$ the row $S^{\b'}_\b(z)$ of size
$\vert n+e_\b\vert=\vert n\vert +1$ which has zeroes everywhere, except in its $\b'$-th block,
namely\footnote{Recall that $n_{\b'}'=n_{\b'}+\delta_{\b\b'}.$}
\begin{equation*}
  S_\b^{\b'}(y)=\left(0 \dots 0\ \left(y^j\right)_{0\leq j<n_{\b'}'}\ 0\dots 0\right).
\end{equation*}%
Notice that with this notation, definition
(\ref{eq:Rtilde}) of $\tilde R_\a(z)$ can be rewritten
as
\begin{equation}\label{RCC}
  \tilde R_\a(z)=\CC_{\psi^{-s}_\a}\left(\sum_{\b'=1}^p S_\b^{\b'}(z)\rhi^t_{\b'}(z)\right).
\end{equation}
It suggests the  introduction of the following
polynomials (in $y$)
\begin{equation}\label{for:S_def}
  \renewcommand{\arraystretch}{1.2}
  S_{\a\b}^{\b'}(y):=\det
  \begin{matrix}{c}
    R_1^0\\ \vdots\\ R_\a^{m_\a-1}\\ S_\b^{\b'}(y)\\ R_{\a+1}^0\\ \vdots\\ R_q^{m_q-1}
  \end{matrix}.
\end{equation}%
Expanding this determinant along its $(m_1+\cdots+m_\a+1)$-th row, which is the (unique) row that contains $y$, it
is clear that if $\b'\neq\b$, then $\deg S_{\a\b}^{\b'}(y)<n'_{\b'}=n_{\b'}$. In view of (\ref{tau_R}) we also
have\footnote{See (\ref{eps2_def}) for the definition of $\e_{\a\b}(m,n)$.}
\begin{equation*}
  S_{\a\b}^{\b}(y)=\e_{\a\b}\,(m,n)\,\tau_{mn}y^{n_\b}+O(y^{n_\b-1}).
\end{equation*}%
Moreover, for any $\a'=1,\dots,q$ and $i=0,\dots,m_{\a'}-1$, we have by linearity of the determinant
\begin{eqnarray}\label{for:orthog}
  \lefteqn{\inn{x^i\psi^{-s}_{\a'}(x)} {\sum_{\b'=1}^pS_{\a\b}^{\b'}(y)\rhi^t_{\b'}(y)}}\nonumber\\
  &=&\inn{x^i\psi^{-s}_{\a'}(x)}{
  \renewcommand{\arraystretch}{1.2}
  \det
  \begin{matrix}{c}
    R_1^0\\ \vdots\\ R_\a^{m_\a-1}\\ \ds\sum_{\b'=1}^p S^{\b'}_\b(y)\rhi^t_{\b'}(y)\\ R_{\a+1}^0\\ \vdots\\ R_q^{m_q-1}
  \end{matrix}
  }\\
  &=&\det
  \renewcommand{\arraystretch}{1.2}
  \begin{matrix}{c}
    R_1^0\\ \vdots\\ R_\a^{m_\a-1}\\
  \renewcommand{\arraystretch}{0.6}
  \Bigl(\inn{x^{i}\psi^{-s}_{\a'}(x)}{y^{j}\rhi^t_{\b'}(y)}\Bigr)_{
    \begin{array}{c}
      \scriptstyle 1\leq \b'<p\\
      \scriptstyle 0\leq j<n'_{\b'}
     \end{array}}
    \\
    R_{\a+1}^0\\ \vdots\\ R_q^{m_q-1}
  \end{matrix}=\det
  \renewcommand{\arraystretch}{1.2}
  \begin{matrix}{c}
    R_1^0\\ \vdots\\ R_\a^{m_\a-1}\\
    R^i_{\a'}\\
    R_{\a+1}^0\\ \vdots\\ R_q^{m_q-1}
  \end{matrix},\nonumber
\end{eqnarray}
which is zero, since the latter matrix has two identical rows ($i<m_{\a'}$).  This shows that
\begin{equation*}
  \frac{\e_{\a\b}(m,n)}{\tau_{mn}}S_{\a\b}^1(y)\,,\ \dots\ ,\,  \frac{\e_{\a\b}(m,n)}{\tau_{mn}}S_{\a\b}^p(y)
\end{equation*}%
are type II mixed mops, normalized with respect to $\rhi^t_\b$. It follows from Proposition \ref{prp:typeII} that
\begin{equation}\label{for:type_II_alt}
  S_{\a\b}^{\b'}(y)=\e_{\a\b}(m,n)\,\tau_{mn}\,\Q{\b,\b'}(y),
\end{equation}%
for any\footnote{The formulas for the different values of $\a$ are all the same, up to a sign, as they amount to
changing the location of a row in the evaluation of a determinant.}  $\a=1,\dots,q$. Since
$\Q{\b}(y)=\Q{\b,1}(y)\rhi^t_1(y)+\cdots+\Q{\b,p}(y)\rhi^t_p(y)$, it follows from (\ref{for:first_eq}), (\ref{RCC}),
(\ref{for:type_II_alt}) and (\ref{tau_Ct}) that
\begin{eqnarray}\label{for:tau_to_cauchy}
  {z^{-m_\a-1}\,\frac{\tau_{m+e_\a,n+e_\b}(s_\a-\zs)}{\tau_{mn}}}
    &=&\frac1{\tau_{mn}}\ \CC_{\psi^{-s}_\a}\left(\sum_{\b'=1}^pS_{\a\b}^{\b'}(z)\rhi^t_{\b'}(z)\right)\nonumber\\
    &=&\e_{\a\b}(m,n)\ \CC_{\psi^{-s}_\a}\left(\sum_{\b'=1}^p\Q{\b,\b'}(z)\rhi^t_{\b'}(z)\right)\nonumber\\
    &=&\e_{\a\b}(m,n)\ \CC_{\psi^{-s}_\a}\Q{\b}(z)
\end{eqnarray}
This finishes the proof. \qed
\end{proof}
Observe our proof shows, as a byproduct, that each
$\Q{\b}(y)$ is expressible naturally as a determinant,
like in the classical case, namely
\begin{equation}\label{eq:byproduct}
  \Q{\b}(y)=\frac{\e_{\a\b}(m,n)}{\tau_{mn}}
    \det
  \begin{matrix}{c}
    R_1^0\\ \vdots\\ R_\a^{m_\a-1}\\ \ds\sum_{\b'=1}^p S^{\b'}_\b(y)\rhi^t_{\b'}(y)\\ R_{\a+1}^0\\ \vdots\\ R_q^{m_q-1}
  \end{matrix}.
\end{equation}%
We now get to the second type of Cauchy transforms which
correspond to the Type I mops $\P{\a}(y)$.
\begin{proposition}\label{prp:cauchy_r}
  For $1\leq \a,\,\a'\leq q$ the Cauchy transforms of $\P{\a'}(y)=\P{\a',1}(y)$
  $\rhi^t_1(y)+\cdots+\P{\a',p}(y)\rhi^t_{p}(y)$ with respect to $\psi^{-s}_{\a}$ can be expressed in terms of tau
  functions as follows:
  \begin{eqnarray}\label{for:cauchy_r}
    \CC_{\psi^{-s}_\a}\P{\a}(z)&=&z^{-m_\a}\,\frac{\tau_{mn}(s_\a-\zs)}{\tau_{mn}},\\
    \CC_{\psi^{-s}_{\a}}\P{\a'}(z)&=&\ve_{\a'\a}(m)\,z^{-1-m_{\a}}\,\frac{\tau_{m+e_{\a}-e_{\a'},n}(s_\a-\zs)}{\tau_{mn}},\qquad
    \a'\neq\a.\label{for:cauchy_r2}
  \end{eqnarray}%
\end{proposition}
\begin{proof}\quad
Up to a relabeling of the indices, the shifted tau
functions in question were already expressed as
polynomials in the previous proof. Let us show how this
leads to a quick proof of (\ref{for:cauchy_r}). Shifting
the $m_\a$ and $n_\b$ indices down by $1$, it follows
from (\ref{for:first_eq}) that
\begin{equation*}
  z^{-m_\a}\tau_{mn}(s_\a-\zs)=\det
  \begin{matrix}{c}
    R_1^0\\ \vdots\\ R_\a^{m_\a-2}\\ \tilde R_\a(z)\\ R_{\a+1}^0\\ \vdots\\ R_q^{m_q-1}
  \end{matrix},
\end{equation*}%
while the orthogonality relations (\ref{for:orthog}) become
\begin{equation}\label{eq:ortho_rel}
  {\inn{x^i\psi^{-s}_{\a'}(x)} {\sum_{\b'=1}^pS_{\a\b}^{\b'}(y)\rhi^t_{\b'}(y)}}
  =\det
  \renewcommand{\arraystretch}{1.2}
  \begin{matrix}{c}
    R_1^0\\ \vdots\\ R_\a^{m_\a-2}\\
    R^i_{\a'}\\
    R_{\a+1}^0\\ \vdots\\ R_q^{m_q-1}
  \end{matrix}
  =\delta_{\a\a'}\delta_{i,m_\a-1}\tau_{mn}.
\end{equation}
Since $\deg S_{\a\b}^{\b'}<n_{\b'}$ for $\b'=1,\dots,p$ this means that the polynomials
\begin{equation*}
  \frac{1}{\tau_{mn}}S_{\a\b}^1(y)\,,\ \dots\ ,\,  \frac{1}{\tau_{mn}}S_{\a\b}^p(y)
\end{equation*}%
are type I mixed mops, normalized with respect to $\psi^{-s}_\a$, so they coincide according to Proposition
\ref{prp:typeI} with the polynomials $\P{\a,1}(y),\dots,\P{\a,p}(y)$. We conclude, as in~(\ref{for:tau_to_cauchy}),
that
\begin{equation*}
  {z^{-m_\a}\,\frac{\tau_{mn}(s_\a-\zs)}{\tau_{mn}}}
    =\frac1{\tau_{mn}}\ \CC_{\psi^{-s}_\a}\left(\sum_{\b'=1}^pS_{\a\b}^{\b'}(z)\rhi^t_{\b'}(z)\right)
    =\CC_{\psi^{-s}_\a}\P{\a}(z).
\end{equation*}%
Similarly, one obtains (\ref{for:cauchy_r2}) from (\ref{for:cauchy}) by shifting $m_\a'$ and $n_\b$ down by $1$;
the sign in this case is determined (for $\a'<\a$) from the right hand side of (\ref{eq:ortho_rel}) now taking the
form
\begin{equation*}
  \det
  \renewcommand{\arraystretch}{1.2}
  \begin{matrix}{c}
    R_1^0\\ \vdots\\ R_{\a'}^{m_{\a'}-2}\\ R^0_{\a'+1}\\ \vdots\\ R_\a^{m_\a-1}\\
    R^i_{\a''}\\
    R_{\a+1}^0\\ \vdots\\ R_q^{m_q-1}
  \end{matrix}
  =\ve_{\a'\a}(m)\,\det
  \begin{matrix}{c}
    R_1^0\\ \vdots\\ R_{\a'}^{m_{\a'}-2}\\ R^i_{\a''}\\R^0_{\a'+1}\\ \vdots\\ R_\a^{m_\a-1}\\
    R_{\a+1}^0\\ \vdots\\ R_q^{m_q-1}
  \end{matrix}
  =\ve_{\a'\a}(n)\,\delta_{\a'\a''}\delta_{i,m_{\a'}-1}\tau_{mn}.
\end{equation*}
\qed
\end{proof}

\goodbreak

\section{Duality}

By interchanging the r\^oles of the weights $\psi_\a^t$
with the weights $\rhi_\b^{-s}$ we obtain \Type
{I}{\rhi^t_\b} mixed mops and \Type{II}{\psi^{-s}_\a}
mixed mops, expressed in terms of tau functions, leading
to a \emph{duality}. As a general rule, in order to
dualize a formula one does the following exchanges
\begin{equation}\label{duality1}
  q\leftrightarrow p,\quad
  m\leftrightarrow n,\quad
  \psi\leftrightarrow\rhi,\quad
  s\leftrightarrow -t,\quad
  x\leftrightarrow y.
\end{equation}%
At the level of the indices, duality amounts to
\begin{equation}\label{duality2}
  \a\leftrightarrow \b,\quad
  i\leftrightarrow j.
\end{equation}%
As for the mixed mops which we have constructed, they
will correspond to new mixed mops for which we will use
the same letter, but adding a star. Thus,
\begin{equation}\label{duality3}
  \P{\a}\leftrightarrow \PS{\b},\quad
  \P{\a,\b}\leftrightarrow \PS{\b,\a},\quad
  \Q{\b}\leftrightarrow \QS{\a},\quad
  \Q{\b,\b'}\leftrightarrow \QS{\a,\a'}.
\end{equation}%
What happens to the tau functions $\tau_{mn}$? To see
this, pick a typical shifted tau function
$\tau_{m+e_\a-e_{\a'},n}(s_\a-\zs)$ and make its
dependence on the weights and on all times explicit,
writing $\tau_{m+e_\a-e_{\a'},n}(s-\zs
e_\a,t;\psi,\rhi)$. According to the above rule it
becomes $\tau_{n+e_\b-e_{\b'},m}(-t-\zs
e_\b,-s;\rhi,\psi)$ which is equal to
$\tau_{m,n+e_\b-e_{\b'}}(s,t+\zs e_\b;\psi,\rhi)$, since
transposing the moment matrix has no effect on the
determinant, while it permutes the indices in the tau
function, it permutes the time-dependence (with signs)
and it permutes the weights. Thus,
\goodbreak
\begin{eqnarray*}
  \tau_{mn}&\leftrightarrow&\tau_{mn}\\
  \tau_{mn}(t_\b-\zs)&\leftrightarrow&\tau_{mn}(s_\a+\zs)\\
  \tau_{m-e_\a,n-e_\b}(t_\b-\zs)&\leftrightarrow&\tau_{m-e_\a,n-e_\b}(s_\a+\zs)\\
  \tau_{m+e_\a-e_{\a'},n}(s_\a-\zs)&\leftrightarrow&\tau_{m,n+e_\b-e_{\b'}}(t_\b+\zs),
\end{eqnarray*}
and so on. Dualizing Propositions \ref{prp:typeII} and \ref{prp:typeI}, we get the following proposition.
\begin{proposition}\label{prp:dual}
   For $\a=1,\dots,q$ and $\b=1,\dots,p$, let
  \begin{eqn}{P*Q*}
    \PS{\b}(x)&:=&\PS{\b,1}(x)\psi^{-s}_1(x)+\cdots+ \PS{\b,q}(x)\psi^{-s}_q(x),\\
    \QS{\a}(x)&:=&\QS{\a,1}(x)\psi^{-s}_1(x)+\cdots+\QS{\a,q}(x)\psi^{-s}_q(x),
  \end{eqn}%
  where $\PS{\b,\a}$ and $\QS{\a,\a'}$ are the polynomials, defined by
  \begin{equation}
    \PS{\b,\a}(z):=\e_{\b\a}(n,m)z^{m_\a-1}\frac{\tau_{m-e_\a,n-e_\b}(s_\a+\zs)}{\tau_{mn}},
  \end{equation}
and
  \begin{equation}
    \begin{array}{rcl}
      \QS{\a,\a}(z)&:=&\ds z^{m_\a}\frac{\tau_{mn}(s_\a+\zs)}{\tau_{mn}}\\
      \QS{\a,\a'}(z)&:=&\ds\ve_{\a\a'}(m)z^{m_{\a'}-1}\frac{\tau_{m+e_\a-e_{\a'},n}(s_{\a'}+\zs)}{\tau_{mn}},\qquad \a'\neq\a.
    \end{array}
  \end{equation}
  Then $\PS{\b,1}(x),\dots,\PS{\b,q}(x)$ are \Type I{\rhi^t_\b} mixed mops, while $\QS{\a1}(x),\dots,\QS{\a,q}(x)$
  are \Type{II}{\psi^{-s}_\a} mixed mops.
\end{proposition}
\qed
\goodbreak

Dualizing Definition (\ref{eq:cauchy_def}) we get the following definition for the dual Cauchy transform: for any
function $F$ and a weight $\rhi$ we put
\begin{equation}\label{eq:cauchy_dual_def}
  \CC^*_\rhi F(z):=\dint_{\R^2}
  \frac{\rhi(y)}{z-y} {F(x)}d\mu(x,y)=
  \oinn{\frac{\rhi(y)}{z-y}}{F(x)}.
\end{equation}%
If we dualize now Propositions \ref{prp:cauchy_q} and \ref{prp:cauchy_r}, then we get the following proposition.
\begin{proposition}
  For $\a=1,\dots,q$ and $\b,\b'=1,\dots,p$, the Cauchy transforms of $\PS{\b'}(x)$ with respect to $\rhi^t_{\b}$, and of
  $\QS{\a}(x)$ with respect to $\rhi^t_\b$ can be expressed in terms of tau functions as follows:
  \begin{eqnarray*}
    \CC^*_{\rhi^t_\b}\PS{\b}(z)&=&z^{-n_\b}\,\frac{\tau_{mn}(t_\b+\zs)}{\tau_{mn}},\\
    \CC^*_{\rhi^t_{\b}}\PS{\b'}(z)&=&\ve_{\b'\b}(n)\,z^{-1-n_{\b}}\,\frac{\tau_{m,n+e_{\b}-e_{\b'}}(t_\b+\zs)}{\tau_{mn}},\qquad
    \b'\neq\b,
  \end{eqnarray*}%
  and
\begin{equation*}
  \CC^*_{\rhi^t_\b}\QS{\a}(z)=\e_{\b\a}(n,m)\,z^{-n_\b-1}\,\frac{\tau_{m+e_\a,n+e_\b}(t_\b+\zs)}{\tau_{mn}}.
\end{equation*}%
\end{proposition}
\qed

\section{The Riemann-Hilbert matrix and the bilinear identity}
\label{sec:RH}
Orthogonal polynomials were shown to be characterized by
a Riemann-Hilbert problem in \cite{FIK} and
\cite{Deift}. This was generalized by Daems and
Kuijlaars to the case of mixed mops. According to
\cite{DK}\footnote{Up to a factor
$\diag(I_p,\,-2\pi\sqrt{-1}I_q)$ which we suppress.} the
corresponding Riemann-Hilbert matrix is given by the
$(p+q)\times(p+q)$ matrix
\begin{equation*}
  Y_{mn}(z):=
  \renewcommand{\arraystretch}{2}
  \begin{matrix}{cc}
    \renewcommand{\arraystretch}{0.6}
    \left(\Q{\b,\b'}\right)_{
    \begin{array}{c}
      \scriptstyle 1\leq \b\leq p\\
      \scriptstyle 1\leq \b'\leq p
     \end{array}}&
    \renewcommand{\arraystretch}{0.6}
    \left(\CC_{\psi^{-s}_\a}\Q{\b}\right)_{
    \begin{array}{c}
      \scriptstyle 1\leq \b\leq p\\
      \scriptstyle 1\leq \a\leq q
     \end{array}}\\
    \renewcommand{\arraystretch}{0.6}
    \left(\P{\a,\b}\right)_{
    \begin{array}{c}
      \scriptstyle 1\leq \a\leq q\\
      \scriptstyle 1\leq \b\leq p
     \end{array}}&
    \renewcommand{\arraystretch}{0.6}
    \left(\CC_{\psi^{-s}_\a}\P{\a'}\right)_{
    \begin{array}{c}
      \scriptstyle 1\leq \a'\leq q\\
      \scriptstyle 1\leq \a\leq q
     \end{array}}
  \end{matrix}=
\end{equation*}%
\begin{equation*}
\begin{matrix}{cc}
  \renewcommand{\arraystretch}{0.6}
  \scriptscriptstyle
    \left(\ve_{\b\b'}(n)\frac{\tau_{m,n+e_\b-e_{\b'}}(t_{\b'}-\zs)} {\tau_{mn}}z^{n_{\b'}+\delta_{\b\b'}-1}
     \right)_{\begin{array}{c}
      \scriptscriptstyle 1\leq \b\leq p\\
      \scriptscriptstyle 1\leq \b'\leq p
     \end{array}}
  &
  \renewcommand{\arraystretch}{0.6}
  \scriptstyle
  \left(\e_{\a\b}(m,n)\frac{\tau_{m+e_\a,n+e_\b}(s_{\a}-\zs)} {\tau_{mn}}z^{-m_{\a}-1}
     \right)_{\begin{array}{c}
      \scriptscriptstyle 1\leq \b\leq p\\
      \scriptscriptstyle 1\leq \a\leq q
     \end{array}}\\
  \renewcommand{\arraystretch}{0.6}
  \scriptstyle
  \left(\e_{\a\b}(m,n)\frac{\tau_{m-e_\a,n-e_\b}(t_{\b}-\zs)} {\tau_{mn}}z^{n_{\b}-1}\right)_{\begin{array}{c}
      \scriptscriptstyle 1\leq \a\leq q\\
      \scriptscriptstyle 1\leq \b\leq p
     \end{array}}
  &
  \renewcommand{\arraystretch}{0.6}
  \scriptstyle
  \left(\ve_{\a'\a}(m)\frac{\tau_{m+e_\a-e_{\a'},n}(s_{\a}-\zs)} {\tau_{mn}}z^{\delta_{\a\a'}-1-m_{\a}}
     \right)_{\begin{array}{c}
      \scriptscriptstyle 1\leq \a'\leq q\\
      \scriptscriptstyle 1\leq \a\leq q
     \end{array}}
\end{matrix}
\end{equation*}%
whose inverse transpose matrix is given by
\begin{equation*}
  Y_{mn}^*(z)=
  \renewcommand{\arraystretch}{2}
  \begin{matrix}{cc}
    \renewcommand{\arraystretch}{0.6}
    \left(\CC^*_{\rhi^t_\b}\PS{\b'}\right)_{
    \begin{array}{c}
      \scriptstyle 1\leq \b'\leq p\\
      \scriptstyle 1\leq \b\leq p
     \end{array}}&
    \renewcommand{\arraystretch}{0.6}
    \left(-\PS{\b,\a}\right)_{
    \begin{array}{c}
      \scriptstyle 1\leq \b\leq p\\
      \scriptstyle 1\leq \a\leq q
     \end{array}}\\
    \renewcommand{\arraystretch}{0.6}
    \left(-\CC^*_{\rhi^t_\b}\QS{\a}\right)_{
    \begin{array}{c}
      \scriptstyle 1\leq \a\leq q\\
      \scriptstyle 1\leq \b\leq p
     \end{array}}&
    \renewcommand{\arraystretch}{0.6}
    \left(\QS{\a,\a'}\right)_{
    \begin{array}{c}
      \scriptstyle 1\leq \a\leq q\\
      \scriptstyle 1\leq \a'\leq q
     \end{array}}
  \end{matrix}=
\end{equation*}%
\begin{equation*}
\begin{matrix}{cc}
  \renewcommand{\arraystretch}{0.6}
  \scriptscriptstyle
    \left(\ve_{\b'\b}(n)\frac{\tau_{m,n+e_\b-e_{\b'}}(t_{\b}+\zs)} {\tau_{mn}}z^{\delta_{\b'\b}-1-n_{\b}}
     \right)_{\begin{array}{c}
      \scriptscriptstyle 1\leq \b'\leq p\\
      \scriptscriptstyle 1\leq \b\leq p
     \end{array}}
  &
  \renewcommand{\arraystretch}{0.6}
  \scriptstyle
  \left(-\e_{\b\a}(n,m)\frac{\tau_{m-e_\a,n-e_\b}(s_{\a}+\zs)} {\tau_{mn}}z^{m_{\a}-1}
     \right)_{\begin{array}{c}
      \scriptscriptstyle 1\leq \b\leq p\\
      \scriptscriptstyle 1\leq \a\leq q
     \end{array}}\\
  \renewcommand{\arraystretch}{0.6}
  \scriptstyle
  \left(-\e_{\b\a}(n,m)\frac{\tau_{m+e_\a,n+e_\b}(t_{\b}+\zs)} {\tau_{mn}}z^{-n_{\b}-1}\right)_{\begin{array}{c}
      \scriptscriptstyle 1\leq \a\leq q\\
      \scriptscriptstyle 1\leq \b\leq p
     \end{array}}
  &
  \renewcommand{\arraystretch}{0.6}
  \scriptstyle
  \left(\ve_{\a\a'}(m)\frac{\tau_{m+e_\a-e_{\a'},n}(s_{\a'}+\zs)} {\tau_{mn}}z^{\delta_{\a\a'}-1+m_{\a'}}
     \right)_{\begin{array}{c}
      \scriptscriptstyle 1\leq \a\leq q\\
      \scriptscriptstyle 1\leq \a'\leq q
     \end{array}}
\end{matrix}
\end{equation*}%
We will obtain bilinear identities for these tau
functions from an identity which is satisfied by the
Riemann-Hilbert matrix and its adjoint. We define the
wave matrix $W_{mn}(z)$ by $Y_{mn}(z)\Delta(z)$, where
$\Delta(z)$ is the diagonal matrix\footnote{Throughout
this section, we set $\xi(t,z):=\sum_1^{\iy}t_kz^k$.}
\begin{equation*}
  \Delta(z):=\diag(\exi {t_1}z,\,\dots, \exi {t_p}z,\ \exi {s_1}z,\ \exi {s_q}z),
\end{equation*}%
with adjoint wave matrix $Y_{mn}^*(z)\Delta^\mi(z)$. In order to make the dependence on the time variables $(s,t)$
explicit, we will write $W_{mn}(z;s,t)$ for $W(z)$ and $W^*_{mn}(z;s,t)$ for $W^*_{mn}(z)$.
\begin{theorem}\label{thm:bilinear}
  The tau functions $\tau_{mn}$ satisfy the following bilinear identities that characterize the tau functions of
  the $(p+q)$-KP hierarchy (see \cite{UT}):
  \begin{equation*}
    \oint_\infty W_{mn}(z;s,t)W^*_{m^*n^*}(z;s^*,t^*)^\top dz=0,
  \end{equation*}%
  which is equivalent to the single identity
  \begin{eqnarray}\label{eq:bilinear_tau}
    \sum_{\b=1}^p\oint_\infty (-1)^{\sigma_\b(n)}\,\tau_{m,n-e_{\b}}(t_{\b}-\zs)\tau_{m^*,n^*+e_{\b}}(t^*_{\b}+\zs)
          \exi {t_{\b}-t_{\b}^*}z\,z^{n_\b-n_\b^*-2}\,dz=\nonumber\\
    \sum_{\a=1}^q\oint_\infty  (-1)^{\sigma_\a(m)}\,\tau_{m+e_\a,n}(s_{\a}-\zs)\tau_{m^*-e_\a,n^*}(s^*_{\a}+\zs)
          \exi {s_{\a}-s_{\a}^*}z\,z^{m_{\a}^*-m_\a-2}\,dz,
  \end{eqnarray}
  where
  \begin{equation}\label{eq:sigma_def}
    \sigma_\a(m)={\sum_{\a'=1}^\a(m_{\a'}-m_{\a'}^*)}\quad\hbox{and}\quad
    \sigma_\b(n)={\sum_{\b'=1}^\b(n_{\b'}-n_{\b'}^*)}.
  \end{equation}%
  and $\vert m^*\vert=\vert n^*\vert+1$ and $\vert m\vert=\vert n\vert-1$.
\end{theorem}%
\begin{proof}\quad
For the entry $(\b',\b'')$ of the product
$$ Y\Delta\transp{(Y^*\Delta^{-1})}=Y\diag(\exi {t_1-t_1^*}z,\,\dots, \exi {t_p-t_p^*}z,\ \exi
  {s_1-s_1^*}z,\,\dots, \exi {s_q-s_q^*}z)\transp{{Y^*}},
$$
we need to prove that
\begin{equation}\label{for:bilin_to_show}
  \sum_{\b=1}^p\oint_\infty  \Q{\b',\b}(z)\,\CC^*_{\rhi^{t^*}_{\b}}\PSS{\b''}(z)\,\exi {t_{\b}-t_{\b}^*}z\,dz=
  \sum_{\a=1}^q\oint_\infty \CC_{\psi^{-s}_\a}\Q{\b'}(z)\,\PSS{\b'',\a}(z)\,\exi {s_{\a}-s_{\a}^*}z\,dz
\end{equation}
where it is understood that all polynomials $P^*$ go with starred times $s^*$ and $t^*$. Also, the integral stands
for (minus) the residue at infinity, and can be computed using the following formal residue identities, with
$f(z)=\sum_{i=0}^\infty a_iz^i$, \renewcommand{\arraystretch}{5}
\begin{eqnarray}\label{eq:rainy}
  \frac1{2\pi\II}\oint_\infty f(z)\,\CC_{\psi} g(z)\,{dz}
  &=&\inn{f(x)\psi(x)}{g(y)},\\
  \frac1{2\pi\II}\oint_\infty \CC^*_\rhi f(z)\, g(z)\,dz
  &=&\inn{f(x)}{\rhi(y)g(y)},\label{eq:rainy2}
\end{eqnarray}%
whose proof we defer until the end. Using this, and Definition (\ref{P*Q*}) of the functions $\PS{\b}(x)$, the left hand
side in (\ref{for:bilin_to_show}) becomes (up to a factor $2\pi\II$)
\begin{eqnarray*}
\sum_{\b=1}^p \inn{\PSS{\b''}(x)}{\Q{\b',\b}(y)\,\rhi^{t^*}_{\b}(y)\,\exi {t_{\b}-t_{\b}^*}y}
  &=&{\sum_{\b=1}^p \inn{\PSS{\b''}(x)}{\Q{\b',\b}(y)\,\rhi^{t}_{\b}(y)}}\\
  &=&\inn{\PSS{\b''}(x)} {\Q{\b'}(y)}.
\end{eqnarray*}
Similarly, the right hand side in (\ref{for:bilin_to_show}) becomes (up to a factor $2\pi\II$)
\begin{eqnarray*}
  \sum_{\a=1}^q \inn{{\PSS{\b'',\a}(x)\,\psi^{-s}_\a}(x)\,\exi {s_{\a}-s_{\a}^*}x} {\Q{\b'}(y)}
  &=&\sum_{\a=1}^q \inn{\PSS{\b'',\a}(x)\,\psi^{-s^*}_\a(x)}{\Q{\b'}(y)}\\
  &=&\inn{\PSS{\b''}(x)} {\Q{\b'}(y)}.
\end{eqnarray*}%
The three other identities are obtained in the same way.

\smallskip

In terms of tau functions, it means that we have shown that for any $m,n,m^*,n^*,\b'$ and $\b''$, with $\vert
m\vert=\vert n\vert$ and $\vert m^*\vert=\vert n^*\vert$ the following bilinear identities hold:
\begin{eqnarray*}
  \sum_{\b=1}^p\oint_\infty  \clubsuit\,\tau_{m,n+e_{\b'}-e_{\b}}(t_{\b}-\zs)\tau_{m^*,n^*+e_{\b}-e_{\b''}}(t^*_{\b}+\zs)
          \exi {t_{\b}-t_{\b}^*}z\,dz=\\
  \sum_{\a=1}^q\oint_\infty  \spadesuit\,\tau_{m+e_\a,n+e_{\b'}}(s_{\a}-\zs)\tau_{m^*-e_\a,n^*-e_{\b''}}(s^*_{\a}+\zs)
          \exi {s_{\a}-s_{\a}^*}z\,dz,
\end{eqnarray*}
where
\begin{eqnarray*}
  \clubsuit&=&\ve_{\b'\b}(n)\ve_{\b''\b}(n^*)z^{n_{\b}-n_{\b}^*-2+\delta_{\b\b'}+\delta_{\b\b''}},\\
  \spadesuit&=&\e_{\a\b'}(m,n)\e_{\b''\a}(n^*,m^*)z^{m^*_\a-m_\a-2}.
\end{eqnarray*}
For different values of $\b'$ and $\b''$ this yields the same identity, up to a relabeling of $n$ and
$n^*$. Namely, replace in the bilinear identity $n+e_{\b'}$ by $n$ and $n^*-e_{\b''}$ by $n^*$ and multiply by
$(-1)^{n_1+\cdots+n_{\b'}}(-1)^{n_1^*+\cdots+n_{\b''}^*}$ to find the following symmetric expression for the
identity, that is independent of $\b'$ and $\b''$:
\begin{eqnarray*}
  \sum_{\b=1}^p\oint_\infty (-1)^{\sigma_\b(n)}\,\tau_{m,n-e_{\b}}(t_{\b}-\zs)\tau_{m^*,n^*+e_{\b}}(t^*_{\b}+\zs)
          \exi {t_{\b}-t_{\b}^*}z\,z^{n_\b-n_\b^*-2}\,dz=\\
  \sum_{\a=1}^q\oint_\infty  (-1)^{\sigma_\a(m)}\,\tau_{m+e_\a,n}(s_{\a}-\zs)\tau_{m^*-e_\a,n^*}(s^*_{\a}+\zs)
          \exi {s_{\a}-s_{\a}^*}z\,z^{m_{\a}^*-m_\a-2}\,dz,
\end{eqnarray*}
where $\sigma_\a(m)$ and $\sigma_\b(n)$
are given by (\ref{eq:sigma_def}).
Notice that, due to the shift, one must
have in this symmetric form that
$\vert m\vert=\vert n\vert-1$ and $\vert m^*\vert=\vert n^*\vert+1$. The other
three identities also yield the above identity, up to relabeling.

\smallskip

Finally, to prove (\ref{eq:rainy}), compute
\begin{eqnarray*}
  \frac1{2\pi\II}\oint_\infty f(z)\inn{\frac{\psi(x)}{z-x}}{g(y)}\,dz=\frac1{2\pi\II}\oint_\infty\sum_{i=0}^\infty
  a_iz^i\sum_{j=0}^\infty \frac1{z^{j+1}}\inn{x^j\psi(x)}{g(y)}\\
  =\sum_{i=0}^\infty a_i\inn{x^i\psi(x)}{g(y)}=\inn{\sum_{i=0}^\infty a_ix^i\,\psi(x)}{g(y)}=\inn{f(x)\psi(x)}{g(y)},
\end{eqnarray*}
and similarly for (\ref{eq:rainy2}), completing the proof.
\qed
\end{proof}

\section{Consequences of the bilinear identities}
In this section we will derive from the bilinear identities (\ref{eq:bilinear_tau}) a series of PDE's for the tau
functions $\tau_{mn}$. In order to keep the formulas transparant we will use the following simplification in the
notation. Recall that we have time variables $s_\a=(s_{\a1},\,s_{\a2},\dots)$ and $t_\b=(t_{\b1},\,t_{\b2},\dots)$,
where $\a=1,\dots,q$ and $\b=1,\dots,p$. In the bilinear identities (\ref{eq:bilinear_tau}) we consider in each
term a shift in $t_\a$, for a single $\a$, or in $s_\b$, for a single $\b$; we will denote this $t_\a$ or $s_\b$ by
$v$ (so $v$ is an infinite vector $v=(v_1,v_2,\dots)$ and we assemble all the other $r:=p+q-1$ series of time
variables in $w=(w_1,w_2,\dots,w_{r})$, where $w_1=(w_{11},w_{12},\dots)$ and so on. Moreover, precisely like in
the bilinear identities we will want to consider an independent collection of all these variables, in fact we will
consider here $(v',w')$ and $(v'',w'')$ besides $(v,w)$. We use the Hirota symbol, which takes in our case the
following form
\begin{equation}\label{eq:Hirota}
  P(\p_v,\p_w)\,F\circ G={P(\p_{v'},\p_{w'})\,F(v+v',w+w')\,G(v-v',w-w')}_{\big\vert_{v'=w'=0}}.
\end{equation}%
The elementary Schur polynomials $S_\ell(v)$ are defined by
\begin{equation}\label{eq:Schur}
  e^{\sum_{k=1}^\infty v_kz^k}=\sum_{k=0}^\infty S_k(v)z^k,
\end{equation}%
for $\ell\geq0$ and $S_{\ell}(v):=0$ otherwise. In particular, if we put degree $v_i:=i$, then
\begin{equation}\label{eq:schur_ex}
  S_0=1,\qquad S_1(v)=v_1,\qquad S_\ell(v)=v_\ell+\hbox{ degree $\ell$ in } v_1,\dots,v_{\ell-1}.
\end{equation}%
We also use the standard notation
\begin{equation*}
  \tilde\p_v=\left(\pp{}{v_1},\,\frac12\pp{}{v_2},\,\frac13\pp{}{v_3},\dots\right).
\end{equation*}%
We first give an identity which will allow us to compute
the formal residues which appear in
(\ref{eq:bilinear_tau}) in terms of derivatives of the
tau function.
\begin{lemma}\label{lma:formal_residue_identity}
  For any $n\in\Z$ we have the following formal residue identity
\begin{eqn}[3]{eq:formal_residue}
  \ds\oint_\infty F(v''+\zs,w'')\,G(v'-\zs,w')\,e^{\sum_{\ell=0}^\infty (v_\ell'-v_\ell'')z^\ell}\,z^n\frac{dz}{2\pi\II}\\
  \ds=\sum_{j\geq0}S_{j-1-n}(-2a)\,S_j(\tilde\partial_v)\,e^{\sum_{\ell=1}^\infty(a_\ell\pp{}{v_\ell}+
    \sum_{\gamma=1}^{r}b_{\gamma \ell} \pp{}{w_{\gamma \ell}})}\,F(v,w)\circ G(v,w),
\end{eqn}
where
\begin{equation*}
\renewcommand{\arraystretch}{1.7}
\begin{array}{cc}
  v'=v-a,\quad v''=v+a,\quad&\quad w_i'=w_i-b_i,\quad w_i''=w_i+b_i,\\
  a=(a_1,a_2,a_3,\dots),\quad&\quad b_i=(b_{i1},\,b_{i2},\,b_{i3},\dots),
\end{array}
\end{equation*}%
for $1\leq i\leq r$.
\end{lemma}
\begin{proof}\quad
The proof is an immediate, but tricky, consequence of Definition (\ref{eq:Schur}) of the Schur functions and of the
following two properties of the Hirota symbol:
\begin{eqnarray*}
  F(v+\zs,w)\,G(v-\zs,w)=\sum_{j=0}^\infty z^{-j}S_j(\tilde\partial_v)\,F\circ G,\\
  F(v+a,w+b)\,G(v-a,w-b)=e^{\sum_{\ell=0}^\infty (a_\ell\pp{}{v_\ell}+\sum_{\gamma=1}^r b_{\gamma
  \ell}\pp{}{w_{\gamma \ell}})}\,F\circ G.
\end{eqnarray*}
\qed
\end{proof}
\begin{proposition}\label{prp:PDE's}
  The bilinear equations imply, upon specialization, that the tau functions $\tau_{mn}$, with $\vert m\vert=\vert
  n\vert$ satisfy the following PDE's expressed in terms of the Hirota symbol:
  \begin{eqnarray}
    \tau_{mn}^2\frac{\p^2}{\p t_{\b,\ell+1}\p t_{\b',1}}\ln\tau_{mn}\label{eq:PDE_in_prop1}
    &=&
    S_{\ell+2\delta_{\b\b'}}(\tilde\partial_{t_{\b}})\tau_{m,n+e_{\b}-e_{\b'}}\circ\tau_{m,n+e_{\b'}-e_{\b}}
    \\
    \tau_{mn}^2\frac{\p^2}{\p s_{\a,\ell+1}\p s_{\a',1}}\ln\tau_{mn}\label{eq:PDE_in_prop2}
    &=&
    S_{\ell+2\delta_{\a\a'}}(\tilde\partial_{s_{\a}})\tau_{m+e_{\a'}-e_{\a},n}\circ\tau_{m+e_{\a}-e_{\a'},n}
    \\
    -\tau_{mn}^2\frac{\p^2}{\p s_{\a,1}\p t_{\b,\ell+1}}\ln\tau_{mn}\label{eq:PDE_in_prop3}
    &=&
    S_{\ell}(\tilde\partial_{t_{\b}})\tau_{m+e_{\a},n+e_\b}\circ\tau_{m-e_{\a},n-e_{\b}}
    \\
    -\tau_{mn}^2\frac{\p^2}{\p t_{\b,1}\p s_{\a,\ell+1}}\ln\tau_{mn}.\label{eq:PDE_in_prop4}
    &=&
    S_{\ell}(\tilde\partial_{s_{\a}})\tau_{m-e_{\a},n-e_\b}\circ\tau_{m+e_{\a},n+e_{\b}}
  \end{eqnarray}
  Equations (\ref{eq:PDE_in_prop1}) resp.\ (\ref{eq:PDE_in_prop2}) for $\b'=\b$ (resp.\ for $\a'=\a$) yield a
  solution to the KP hierarchy in $t_\b$ (resp.\ in $s_\a$), while for $\b'\neq\b$ and $\a'\neq\a$,
  (\ref{eq:PDE_in_prop1}) --- (\ref{eq:PDE_in_prop4}) yields
  \begin{eqnarray}
    \frac{\p^2}{\p t_{\b,1}\p t_{\b',1}}\ln\tau_{mn}&=&\frac{\tau_{m,n+e_{\b}-e_{\b'}}\tau_{m,n+e_{\b'}-e_{\b}}}
      {\tau_{mn}^2}\label{eq:PDE_in_prop5}\\
    \frac{\p^2}{\p s_{\a,1}\p s_{\a',1}}\ln\tau_{mn}&=&\frac{\tau_{m+e_{\a'}-e_{\a},n}\tau_{m+e_{\a}-e_{\a'},n}}
      {\tau_{mn}^2}\label{eq:PDE_in_prop6}\\
    \frac{\p^2}{\p s_{\a,1}\p t_{\b,1}}\ln\tau_{mn}&=&-\frac{\tau_{m+e_\a,n+e_{\b}}\tau_{m-e_\a,n-e_{\b}}}
      {\tau_{mn}^2}\label{eq:PDE_in_prop7}\\
    \frac{\p}{\p t_{\b,1}}\ln\frac{\tau_{m,n+e_\b-e_{\b'}}}{\tau_{m,n+e_{\b'}-e_{\b}}}&=&
      \frac{\frac{\p^2}{\p t_{\b,2}\p t_{\b',1}}\ln\tau_{mn}}{\frac{\p^2}{\p t_{\b,1}\p
      t_{\b',1}}\ln\tau_{mn}}\label{eq:PDE_in_prop8}\\
    \frac{\p}{\p s_{\a,1}}\ln\frac{\tau_{m-e_\a+e_{\a'},n}}{\tau_{m-e_{\a'}+e_{\a},n}}&=&
      \frac{\frac{\p^2}{\p s_{\a,2}\p s_{\a',1}}\ln\tau_{mn}}{\frac{\p^2}{\p s_{\a,1}\p
      t_{\a',1}}\ln\tau_{mn}}\label{eq:PDE_in_prop9}\\
    \frac{\p}{\p t_{\b,1}}\ln\frac{\tau_{m+e_\a,n+e_\b}}{\tau_{m-e_\a,n-e_{\b}}}&=&
      \frac{\frac{\p^2}{\p t_{\b,2}\p s_{\a,1}}\ln\tau_{mn}}{\frac{\p^2}{\p t_{\b,1}\p
      s_{\a,1}}\ln\tau_{mn}}\label{eq:PDE_in_prop10}\\
    \frac{\p}{\p s_{\a,1}}\ln\frac{\tau_{m-e_\a,n-e_\b}}{\tau_{m+e_\a,n+e_{\b}}}&=&
      \frac{\frac{\p^2}{\p s_{\a,2}\p t_{\b,1}}\ln\tau_{mn}}{\frac{\p^2}{\p s_{\a,1}\p
      t_{\b,1}}\ln\tau_{mn}}\label{eq:PDE_in_prop11}.
\end{eqnarray}
  It leads to the following ${p+q}\choose2$ PDE's for $\ln\tau_{mn}$ involving not just one $s_\a$ or $t_\b$, but a
  few of them
\begin{eqnarray}
  \pp{}{t_{\b',1}}\left(\frac{\frac{\p^2}{\p t_{\b,2}\p t_{\b',1}}\ln\tau_{mn}}{\frac{\p^2}{\p t_{\b,1}\p
      t_{\b',1}}\ln\tau_{mn}}\right)+
  \pp{}{t_{\b,1}}\left(\frac{\frac{\p^2}{\p t_{\b',2}\p t_{\b,1}}\ln\tau_{mn}}{\frac{\p^2}{\p t_{\b',1}\p
      t_{\b,1}}\ln\tau_{mn}}\right)&=&0,\label{eq:PDE_in_prop12}\\
  \pp{}{s_{\a',1}}\left(\frac{\frac{\p^2}{\p s_{\a,2}\p s_{\a',1}}\ln\tau_{mn}}{\frac{\p^2}{\p s_{\a,1}\p
      s_{\a',1}}\ln\tau_{mn}}\right)+
  \pp{}{s_{\a,1}}\left(\frac{\frac{\p^2}{\p s_{\a',2}\p s_{\a,1}}\ln\tau_{mn}}{\frac{\p^2}{\p s_{\a',1}\p
      s_{\a,1}}\ln\tau_{mn}}\right)&=&0,\label{eq:PDE_in_prop13}\\
  \pp{}{s_{\a,1}}\left(\frac{\frac{\p^2}{\p t_{\b,2}\p s_{\a,1}}\ln\tau_{mn}}{\frac{\p^2}{\p t_{\b,1}\p
      s_{\a,1}}\ln\tau_{mn}}\right)  +
  \pp{}{t_{\b,1}}\left(\frac{\frac{\p^2}{\p s_{\a,2}\p t_{\b,1}}\ln\tau_{mn}}{\frac{\p^2}{\p s_{\a,1}\p
      t_{\b,1}}\ln\tau_{mn}}\right)&=&0.\label{eq:PDE_in_prop14}
\end{eqnarray}
\end{proposition}
\begin{proof}\quad
Let us denote for $a=(a_1,\dots,a_q)$ and $b=(b_1,\dots,b_q)$ by $\Omega(a,b)$ the differential operator
\begin{equation}\label{eq:Omega_def}
  \Omega(a,b):={\sum_{\ell=1}^\infty\left(\sum_{\a'=1}^qa_{\a'\ell}\pp{}{s_{\a'\ell}}+\sum_{\b'=1}^pb_{\b'\ell}\pp{}{t_{\b'\ell}}\right)}.
\end{equation}%
Using Lemma (\ref{lma:formal_residue_identity}), rewrite the bilinear identity\footnote{Recall that in this form of
the bilinear identity $\vert m^*\vert=\vert n^*\vert+1$ and $\vert m\vert=\vert n\vert-1$.}
(\ref{eq:bilinear_tau}):
\begin{eqn}{eq:bilinear_again}
    &&\ds\sum_{\b=1}^p(-1)^{\sigma_\b(n)}\,\sum_{k=0}^\infty S_{n_\b^*-n_\b+1+k}(-2b_{\b})S_k(\tilde\partial_{t_\b})
       e^{\Omega(a,b)}\tau_{m^*,n^*+e_{\b}}\circ\tau_{m,n-e_{\b}}\\
    &-&\ds\sum_{\a=1}^q(-1)^{\sigma_\a(m)}\,\sum_{k=0}^\infty S_{m_\a-m_\a^*+1+k}(-2a_\a)S_k(\tilde\partial_{s_\a})
       e^{\Omega(a,b)}\tau_{m^*-e_\a,n^*}\circ\tau_{m+e_\a,n}=0.
  \end{eqn}%
Note that all infinite vectors $a_\a$ and $b_\b$ can be chosen completely arbitrary. We set all components of $a$
and $b$ equal to zero, \emph{except} $b_{\b,\ell+1}=B\neq0$ (for some fixed $\b$ and $\ell$), and we set $m^*=m$ and
$n^*-n=-2e_{\b'}$ (for some fixed $\b'$). Then only the first term in (\ref{eq:bilinear_again}) survives, the
signs $\sigma_\beta(n)$ are all $1$ (see (\ref{eq:sigma_def})) and, in view of (\ref{eq:schur_ex}), the identity
(\ref{eq:bilinear_again}) becomes
\begin{eqnarray*}
  0&=&\sum_{\b''=1}^p \sum_{k=0}^\infty S_{1+k-2\delta_{\b'\b''}}(-2b_{\b''})S_k(\tilde\partial_{t_{\b''}})
       e^{B\pp{}{t_{\b,\ell+1}}}\tau_{m,n+e_{\b''}-2e_{\b'}}\circ\tau_{m,n-e_{\b''}}\\
   &=&B\left(-2S_{\ell+2\delta_{\b\b'}}(\tilde\partial_{t_{\b}})\tau_{m,n+e_{\b}-2e_{\b'}}\circ\tau_{m,n-e_{\b}}+\frac{\p^2}{\p
         t_{\b',1}\p t_{\b,\ell+1}}\tau_{m,n-e_{\b'}}\circ\tau_{m,n-e_{\b'}}\right)+O(B^2).
\end{eqnarray*}
Expressing that the coefficient of $B$ in this expression must vanish we get (\ref{eq:PDE_in_prop1}), upon relabeling
$n-e_{\b'}\mapsto n$ and upon using the following property of the Hirota symbol, valid for $f$ depending on (time-)
variables $s$ and $t$:
\begin{equation}\label{eq:hirota_ln}
  \frac{\p^2}{\p t\p s}\,F\circ F=2F^2\frac{\p^2}{\p t\p s}\ln F.
\end{equation}%
(\ref{eq:PDE_in_prop2}) follows from (\ref{eq:PDE_in_prop1}) by duality, using $P(-\p_s)\,F\circ G=P(\p_s)\,G\circ
F$. In order to obtain (\ref{eq:PDE_in_prop3}) we consider again (\ref{eq:bilinear_again}), with
$b_{\b,\ell+1}=B\neq0$ and all other components of $a$ and $b$ equal to zero, but we set now $n^*=n$ and
$m-m^*=-2e_\a$. Then (\ref{eq:bilinear_again}) becomes
\begin{eqnarray*}
  0&=&\sum_{k=0}^\infty S_{k+1}(-2b_\b)S_k(\tilde\p_\b)\,e^{B\pp{}{t_{\b,\ell+1}}}\,\tau_{m+2e_\a,n+e_\b}\circ \tau_{m,n-e_\b}\\
   &&   -\sum_{k=0}^\infty S_{k-1}(0)S_k(\tilde\p_{s_\a})\,e^{B\pp{}{t_{\b,\ell+1}}}\,\tau_{m+e_\a,n}\circ  \tau_{m+e_\a,n}\\
   &=&-B\left(2S_\ell(\tilde\p_{t_\b})\tau_{m+2e_\a,n+e_\b}\circ\tau_{m,n-e_\b}+\frac{\p^2}{\p s_{\a,1}\p t_{\b,\ell+1}}
        \tau_{m+e_\a,n}\circ  \tau_{m+e_\a,n}\right)+O(B^2).
\end{eqnarray*}
The nullity of the coefficient of $B$ in this expression, rewritten by using (\ref{eq:hirota_ln}), leads at once to
(\ref{eq:PDE_in_prop3}), upon doing the relabeling $m+e_\a\mapsto m$. From it, (\ref{eq:PDE_in_prop4}) follows by
duality. Equations (\ref{eq:PDE_in_prop5}) --- (\ref{eq:PDE_in_prop7}) follow from (\ref{eq:PDE_in_prop1}) ---
(\ref{eq:PDE_in_prop3}) by setting $\b'\neq\b,\ \a'\neq\a$ and $\ell=0$. Equations (\ref{eq:PDE_in_prop8}) ---
(\ref{eq:PDE_in_prop11}) follow from (\ref{eq:PDE_in_prop1}) --- (\ref{eq:PDE_in_prop4}) by setting $\b'\neq\b,$
$\a'\neq\a$ and forming in each equation the ratio of the cases $\ell=0$ and $\ell=1$, and using the following property
of the Hirota symbol, valid for $F$ and $G$ depending on a (time-) variable $t$:
\begin{equation*}
  \frac{\p}{\p t}\,F\circ G=FG\,\frac{\p}{\p t}\left(\ln \frac FG\right).
\end{equation*}%
Equations (\ref{eq:PDE_in_prop12}) --- (\ref{eq:PDE_in_prop14}) are just respectively the compatibility equations
between (\ref{eq:PDE_in_prop8}) and $(\ref{eq:PDE_in_prop8})_{\b\leftrightarrow\b'}$, between
(\ref{eq:PDE_in_prop9}) and $(\ref{eq:PDE_in_prop9})_{\a\leftrightarrow\a'}$, and between between
(\ref{eq:PDE_in_prop10}) and (\ref{eq:PDE_in_prop11}).
\qed
\end{proof}
\begin{cor}
  The tau functions $\tau_{mn}$ and the polynomials $\PSS{\b''}(x)=\PSS{\b''}(x,s^*,t^*),$
  $\Q{\b'}(y)=\Q{\b'}(y,s,t)$ appearing in $Y^*$ and $Y$ respectively, satisfy the following $4$ formal series
  identities ($\delta_{\b'\b''}(n,n^*)=(-1)^{n_1+n_2+\cdots+n_{\b'}+n_1^*+n_2^*+\cdots+n_{\b''}^*}$):
  \begin{eqnarray*}
    0&=&\delta_{\b'\b''}(n,n^*)\tau_{mn}(s,t)\tau_{m^*n^*}(s^*,t^*)\inn{\PSS{\b''}(x)}{\Q{\b'}(y)}_{\left|
      \begin{array}{l}
        m^*\mapsto m\\
        n\mapsto n-e_{\b'}+e_{\hat\b},\ n^*\mapsto n+e_{\b''}-e_{\hat\b}\\
        s_\a^*\mapsto s_\a,\ t_\b\mapsto t_\b-b_\b,\ t_\b^*\mapsto t_\b+b_\b
      \end{array}
      \right.}\\
    &=&\sum_{\b=1}^p\sum_{\ell=0}^\infty b_{\b,\ell+1}\left(
      \begin{array}{c}
        \ds-2S_{\ell+2\delta_{\b\hat\b}} (\tilde\p_{t_\b})\tau_{m,n+e_\b-e_{\hat\b}}\circ \tau_{m,n-e_\b+e_{\hat\b}}\\
        \ds+\frac{\p^2}{\p t_{\hat\b,1}\p t_{\b,\ell+1}}\tau_{mn}\circ\tau_{mn}
      \end{array}
      \right)
        +O(b^2),
  \end{eqnarray*}
  \begin{eqnarray*}
    0&=&\delta_{\b'\b''}(n,n^*)\tau_{mn}(s,t)\tau_{m^*n^*}(s^*,t^*)\inn{\PSS{\b''}(x)}{\Q{\b'}(y)}_{\left|
      \begin{array}{l}
        m\mapsto m-e_{\hat\a},\ m^*\mapsto m+e_{\hat\a}\\
        n\mapsto n-e_{\b'},\ n^*\mapsto n+e_{\b''}\\
        s_\a\mapsto s_\a-a_\a,\ s_\a^*\mapsto s_\a+a_\a,\ t_\b^*\mapsto t_\b
      \end{array}
      \right.}\\
    &=&\sum_{\a=1}^q\sum_{\ell=0}^\infty a_{\a,\ell+1}\left(
      \begin{array}{c}
        \ds2S_{\ell+2\delta_{\a\hat\a}} (\tilde\p_{s_\a})\tau_{m+e_{\hat\a}-e_\a,n}\circ \tau_{m-e_{\hat\a}+e_\a,n}\\
        \ds-\frac{\p^2}{\p s_{\hat\a,1}\p s_{\a,\ell+1}}\tau_{mn}\circ\tau_{mn}
      \end{array}
      \right)
        +O(a^2),
  \end{eqnarray*}
  \begin{eqnarray*}
    \lefteqn{\delta_{\b'\b''}(n,n^*)\tau_{mn}(s,t)\tau_{m^*n^*}(s^*,t^*)\inn{\PSS{\b''}(x)}{\Q{\b'}(y)}_{\left|
      \begin{array}{l}
        m\mapsto m-e_{\hat\a},\ m^*\mapsto m+e_{\hat\a}\\
        n\mapsto n-e_{\b'},\ n^*\mapsto n+e_{\b''}\\
        s_\a^*\mapsto s_\a,\ t_\b\mapsto t_\b-b_\b,\ t_\b^*\mapsto t_\b+b_\b
      \end{array}
      \right.}}\hskip 1.85cm\\
    &=&-2\sum_{\b=1}^p\sum_{\ell=0}^\infty b_{\b,\ell+1} S_{\ell} (\tilde\p_{t_\b})\tau_{m+e_{\hat\a},n+e_\b}\circ
               \tau_{m-e_{\hat\a},n-e_\b}+O(b^2)
  \end{eqnarray*}
and
\begin{equation*}
      =\sum_{\b=1}^p\sum_{\ell=0}^\infty b_{\b,\ell+1}\frac{\p^2}{\p s_{\hat\a,1}\p t_{\b,\ell+1}}\tau_{mn}\circ\tau_{mn}+O(b^2),
\end{equation*}%
  \begin{eqnarray*}
    \lefteqn{\delta_{\b'\b''}(n,n^*)\tau_{mn}(s,t)\tau_{m^*n^*}(s^*,t^*)\inn{\PSS{\b''}(x)}{\Q{\b'}(y)}_{\left|
      \begin{array}{l}
        m^*\mapsto m\\
        n\mapsto n-e_{\b'}+e_{\hat\b},\ n^*\mapsto n+e_{\b''}-e_{\hat\b}\\
        s_\a\mapsto s_\a-a_\a,\ s_\a^*\mapsto s_\a+a_\a,\ t_\b^*\mapsto t_\b
      \end{array}
      \right.}}\\
      &=&\sum_{\a=1}^q\sum_{\ell=0}^\infty a_{\a,\ell+1}\frac{\p^2}{\p t_{\hat\b,1}\p
            s_{\a,\ell+1}}\tau_{mn}\circ\tau_{mn}+O(a^2),\hphantom{ddddddddddd}
  \end{eqnarray*}
and
\begin{equation*}
    =-2\sum_{\a=1}^q\sum_{\ell=0}^\infty a_{\a,\ell+1} S_{\ell} (\tilde\p_{s_\a})\tau_{m-e_{\a},n-e_{\hat\b}}\circ
               \tau_{m+e_{\a},n+e_{\b}}+O(a^2).
\end{equation*}%
\end{cor}
\begin{proof}
>From the proof of Theorem \ref{thm:bilinear} and (\ref{eq:formal_residue}) it follows that
  \begin{eqnarray*}
    &&\delta_{\b'\b''}(n,n^*)\tau_{mn}(s,t)\tau_{m^*n^*}(s^*,t^*)\inn{\PSS{\b''}(x)}{\Q{\b'}(y)}_{\left|
      \begin{array}{l}
        n\mapsto n-e_{\b'},\ n^*\mapsto n^*+e_{\b''}\\
        s_\a\mapsto s_\a-a_\a,\ s_\a^*\mapsto s_\a+a_\a\\
        t_\b\mapsto t_\b-b_\b,\ t_\b^*\mapsto t_\b+b_\b
      \end{array}
      \right.}\\
    &=&\sum_{\b=1}^p(-1)^{\s_\b(n)}\sum_{k=0}^\infty S_{n_{\b}^*-n_\b+1+k}(-2b_\b)S_k(\tilde\p_{t_\b}) e^{\Omega(a,b)}
        \tau_{m^*,n^*+e_\b}\circ \tau_{m,n-e_\b}
  \end{eqnarray*}
and
\begin{equation*}
  =\sum_{\a=1}^q(-1)^{\s_\a(m)}\sum_{k=0}^\infty S_{m_{\a}-m_{\a}^*+1+k}(-2a_\a)S_k(\tilde\p_{s_\a}) e^{\Omega(a,b)}
        \tau_{m^*-e_\a,n^*}\circ \tau_{m+e_\a,n}
\end{equation*}%
and so if we just follow the 4 specializations leading to (\ref{eq:PDE_in_prop1}) -- (\ref{eq:PDE_in_prop4}), in
order, we find the 4 equations of the corollary, in their given order.
\qed
\end{proof}

\section{Examples}
\label{sec:examples}
\subsection{Biorthogonal polynomials ($p=q=1$)}
Given the (not necessarily symmetric) inner product with regard to the weight $\rho(x,y)$ on $\R^2$,
$$
  \inn{f(x)}{g(y)}:=\int\!\!\!\!\int_{\R^2}f(x)g(y)\rho(x,y)\,dx\, dy
$$
and the deformed weight
$$
  \rho_{t,s}(x,y)=e^{\sum_1^{\iy}(t_ky^k- s_k x^k)}\rho (x,y).
$$
Setting $p=q=1,~m=m_1,~n=n_1$, with $m=n$, implies that the indices $m,n$ in $\tau_{mn}$ can be replaced by one
single index; namely, set $\tau_n:=\tau_{mn}$, where
$$
 \tau_n(t,s)=\det \Bigl(\!\inn{x^ie^{-\sum_1^{\iy} s_kx^k }}{y^je^{\sum_1^{\iy}t_ky^k}}\Bigr)_{0\leq i,j\leq n-1 } .
$$
Moreover, set $\psi_1=\varphi_1=1$ and define the monic polynomials $p_n^{(1)}(y):=p_n^{(1)}(t,s;y)$ and
$p_n^{(2)}(x):=p_n^{(2)}(t,s;x)$ (with $h_{n-1}^{-1}$ the leading coefficient of $\PS{1,1}(x)$) by
\begin{eqnarray*}
  p_n^{(1)}(y)&:=& Q_{mn}^{(1,1)}(y)=y^n+\cdots\\
  h_{n-1}^{-1}p_{n-1}^{(2)}(x)&:=&\PS{1,1}(x)=h_{n-1}^{-1}x^{n-1}+\cdots
\end{eqnarray*}
The orthogonality conditions (\ref{perp1}) and (\ref{perp2}) imply
\begin{eqnarray*}
  \inn{x^ie^{-\sum_1^{\iy} s_kx^k}}{p^{(1)}_n(y)e^{\sum_1^{\iy} t_ky^k}}&=&0~~\mbox{for}~~0\leq i\leq n-1 \\ \\
  \inn{h_{n }^{-1}p_{n }^{(2)}(x)e^{-\sum_1^{\iy} s_kx^k}}{y^je^{ \sum_1^{\iy} t_ky^k}} &=&0~~\mbox{for}~~0\leq j\leq n-1\\
  &=&1~~\mbox{for}~~ j=n.
\end{eqnarray*}
for all $n\geq 0$, from which the bi-orthogonality can be deduced\footnote{It turns out that
$h_n=\tau_{n+1}(t,s)/\tau_n(t,s)$.}
$$
  \int\!\!\!\!\int_{\R^2}p_n^{(2)}(x) p_m^{(1)}(y)  \rho_{t,s} (x,y)dxdy =\delta_{nm}h_n.
$$
>From (\ref{introI}), (\ref{introII}), (\ref{introIII}) and (\ref{introIV}) and from $h_n=\tau_{n+1}/ \tau_n$, it
follows that
\begin{eqnarray}\label{7.taushifts}
  z^n{ \tau_n(t-[z^{-1}],s)\over\tau_n(t,s)}&=&p^{(1)}_n(z)\nonumber\\
  z^{n } \frac{\tau_{n }(t,s+[z^{-1}])} {\tau_{n}(t,s)}&=&p^{(2)}_{n }(z^{})\nonumber\\
  z^{-n-1}\frac{\tau_{n+1}(t+[z^{-1}],s)}{\tau_n(t,s)}&=&\int\!\!\!\!\int_{\R^2}\frac{p^{(2)}_{n}(x)}{z-y}
               \rho_{t,s}(x,y)dxdy\nonumber\\
  z^{-n-1}\frac{\tau_{n+1}(t,s-[z^{-1}])}{\tau_{n}(t,s)}
  &=& \int\!\!\!\! \int_{\R^2}\frac{p^{(1)}_n(y)}{z-x}
  \rho_{t,s}(x,y)dxdy.
\end{eqnarray}
and from (\ref{eq:bilinear_tau}), the bilinear identity becomes
\begin{eqnarray*}
  \oint_{z=\iy}\tau_{n-1}(t-[z^{-1}],s)\,\tau_{m+1}(t'+[z^{-1}],s')\,e^{\sum_1^{\iy}(t_i-t'_i)z^i} z^{n-m-2}dz\\
  =\oint_{z=\iy}\tau_{n}(t,s-[z^{-1}])\,\tau_m(t',s'+[z^{-1}])\,e^{\sum_1^{\iy}(s_i-s'_i)z^{i}}z^{m-n }dz,
\end{eqnarray*}
which characterizes the $\tau$-functions for the
\emph{2-component KP hierarchy}. Equations
(\ref{7.taushifts}) and the bilinear identity were
obtained in \cite{AvM0}. Indicating the dependence on
$t,s$ in the polynomials, the following inner product
can be computed in two different ways, leading
to\footnote{This integral is $\neq 0$, unless $t=t'$}
\begin{eqnarray}
   &&\tau_n(t,\!s)\tau_{n+1}(t',\!s')\!\!\left.\int\!\!\!\!\int_{\R^2}\!\!\! dx dy ~p_{n+1}^{(2)}(t',s';x) p_{n}^{(1)}(t,s;y)
        e^{\sum_1^{\iy} (t_ky^k-s_k'x^k)}\rho(x,\!y)
   {\begin{array}{l} \\ \\ \\ \end{array}}\right|_{\!\!\!\begin{array}{l}{t\mapsto t-a}\\ {t'\!\mapsto t'\!+\!a}\\{s'=s}
     \end{array}}
   \nonumber\\
   & & =\left(\sum_{j=0}^{\iy} -2a_{j+1}{  S}_j(\tilde\p_t)\tau_{n+2}\circ \tau_n +O(a^2)\right)\nonumber\\
   & & =\left(\sum_{k=1}^{\iy} a_k\frac{\p^2}{\p t_k\p s_1}\tau_{n+1}\circ\tau_{n+1} +O(a^2)\right).
\end{eqnarray}
Identifying the coefficients of $a_{j+1}$ in both expressions and shifting $n\mapsto n-1$ yield a first identity;
then redoing the calculation above for ${s\mapsto s-b},~{s'\mapsto s'+b}$ and $t'=t$ leads to a second one. All in
all we find
%
\begin{eqnarray*}
  {S}_j(\tilde\p_t)\tau_{n+1}\circ\tau_{n-1}&=& -\tau^2_{n} \frac{\p^2}{\p s_1\p t_{j+1}}\ln\tau_{n},\\ \\
  {S}_j(\tilde\p_s)\tau_{n-1 }\circ\tau_{n+1 }&=&-\tau^2_{n}\frac{\p ^2}{\p t_1\p s_{j+1}}\ln\tau_{n}.
\end{eqnarray*}
Specializing the identity (\ref{eq:PDE_in_prop14}) leads to an identity, which can be expressed as a sum of two
Wronskians\footnote{in terms of the Wronskian $\{ f,g \}_t=\frac{\p f}{\p t} g-f \frac{\p g}{\p t}.$} and which
involves a single tau function:
\begin{equation}
  \left\{\frac{\p^2 \ln \tau_{n}}{\p t_1 \p s_2}, \frac{\p^2 \ln \tau_{n}}{\p t_1 \p s_1}
  \right\}_{t_1} +\left\{\frac{\p^2 \ln \tau_{n}}{\p s_1 \p t_2}, \frac{\p^2 \ln \tau_{n}}{\p t_1 \p s_1}
  \right\}_{s_1} =0.
\end{equation}
The computation (2.2) was at the origin of the crucial
argument (Theorem \ref{thm:bilinear}) in this paper. It
illustrates in a simple way what is being done in this
paper. These equations are used, when computing the PDE
for the Dyson, Airy and Sine processes
(\cite{MR2150191}).

\subsection{Orthogonal polynomials}

Given a weight $\rho(z)$ on $\R$, the symmetric inner product
$$
  \inn{f(x)}{g(x)}=\int_{\R}f(x)g(x)\rho(x)\,dx,
$$
and the formal deformation by means of an exponential $\rho_t(x):=\rho(x)e^{\sum_{1}^{\iy}t_kz^k}.$ This is a
special case of the previous example, where the deformation only depends on $t-s$; thus $t-s$ can be replaced by
$t$. Then $\tau_n(t)$ is the determinant of the moment matrix depending on $t=(t_1,t_2,\ldots)$,
$$
  \tau_n(t):=\det\Bigl(\int_{\R}z^{i+j}e^{\sum_{1}^{\iy}t_kz^k}\rho_t(z)dz\Bigr)_{0\leq i,j\leq n-1}.
$$
Then, from (\ref{7.taushifts}), it follows at once that the orthogonal polynomials $p_n(x):=p_n(t;x)$ are given by
\begin{eqnarray*}
  z^n{\tau_n(t-[z^{-1}])\over\tau_n(t)}&=& p_n(z)\\
  z^{ -n-1}\frac{\tau_{n+1}(t+[z^{-1}] )}{\tau_{n }(t )}&=& \int_{\R} \frac{p _n(x)}{z- x}\rho_{t}(x )dx.
\end{eqnarray*}
Moreover, the integral below can be computed in two different ways: on the one hand, it is automatically zero,
because $p_n(z)$ is perpendicular to any polynomial of lower degree; on the other hand, for $t$ and $t'$ close to
each other, the integral can also be developed, using the technique of Proposition \ref{prp:PDE's}, in $t'-t=2y$,
yielding the following formula
\begin{eqnarray*}
  0&=&\tau_n(t)\tau_n(t')\int_{\R} p_n(t;z)p_{n-1}(t',z)
 \rho_t(z)dz\Bigr|_{
   \begin{array}{l}
      t\mapsto t-y\\
      t'\mapsto t+y
   \end{array}}\\
  & =&\sum_3^{\iy} y_k\left(\frac{\p^2}{\p t_1\p t_k}-2{S}_{k+1}\Bigl(\tilde \p_t\Bigr)\right)\tau_n\circ\tau_n+O(y^2),
\end{eqnarray*}
showing that $\tau_n(t)$ satisfies the {\em KP hierarchy}.

\subsection{Orthogonal polynomials on the circle}

Consider the inner product on the circle between analytic functions on $S^{1}$:
$$
  \inn{f(z)}{g(z)}=\oint_{S^{1}} \frac{dz}{2\pi \II z}f(z^{-1})g(z)
$$
and the determinant of moment matrices
\begin{eqnarray*}
  \tau_n(t,s)
    &:=&\det\Bigl(\!\inn{z^ke^{-\sum_1^{\iy}s_i z^i}}{z^\ell e^{\sum_1^{\iy}t_iz^i } } \Bigr)_{0\leq k,\ell\leq n-1 }\\
    &=&\det\left(\oint_{S^{1}}\frac{dz}{2\pi\II z}z^{-k+\ell}e^{\sum_1^{\iy}(t_iz^i-s_iz^{-i})}\right)_{0\leq k,\ell\leq n-1}
\end{eqnarray*}
Then it follows that
\begin{eqnarray*}
  z^n{ \tau_n(t-[z^{-1}],s)\over\tau_n(t,s)}&=&p^{(1)}_n(z)\nonumber\\
  z^n\frac{\tau_n(t,s+[z^{-1}])}{\tau_n(t,s)}&=&p_n^{(2)}( z)\nonumber\\
  z^{- n-1}\frac{\tau_{n+1 }(t+[z^{-1}],s)}{\tau_{n}(t,s)}
      &=&\oint_{S^1}\frac{du}{2\pi\II u}\frac{p^{(2)}_{n}(u^{-1})}{z-u}e^{\sum_1^{\iy}(t_iu^i-s_iu^{-i})}\nonumber\\
  z^{-n-1}\frac{\tau_{n+1}(t,s-[z^{-1}])}{\tau_{n}(t,s)}
      &=& \oint_{S^1}\frac{du}{2\pi i u}\frac{p^{(1)}_n(u)}{z-u^{-1}}e^{\sum_1^{\iy}(t_iu^i-s_iu^{-i})},
\end{eqnarray*}
with $p_n^{(1)}( z)$ and $p_m^{(2)}(z^{-1})$ monic orthogonal polynomials on the circle:
$$
  \oint_{S^{1}} \frac{dz}{2\pi i z}p_n^{(1)}( z)p_m^{(2)}(z^{-1})=\delta_{nm}h_n,~~\mbox{with}~
  h_n=\frac{\tau_{n+1}}{\tau_n}.
$$
The nature of the inner product implies some extra-relationship between the orthogonal polynomials
\begin{eqnarray*}
  p^{(1)}_{n+1}(z)-zp_n^{(1)}(z)&=&p^{(1)}_{n+1}(0)z^np_n^{(2)}(z^{-1})\\
  p^{(2)}_{n+1}(z)-zp_n^{(2)}(z)&=&p^{(2)}_{n+1}(0)z^np_n^{(1)}(z^{-1}).
\end{eqnarray*}
leading to (in the notation of footnote 8)
\begin{eqnarray*}
  &&\left(\frac{h_{n}}{h_{m+1}}\right)^{2}\left(1-\frac{h_{n+1}}{h_{n}}\right)\left(1-\frac{h_{m+1}}{h_{m}}\right)\\
  &&  \\
  &&~~~~=\frac{1}{\tau_{m+2}^2\tau_n^2}\left(S_{n-m}(\tilde\p_{t})\tau_{m+2}\circ\tau_{n}\right)~.~
    \left(S_{n-m}(-\tilde\p_{s})\tau_{m+2}\circ\tau_{n}\right) ~.
\end{eqnarray*}
In particular, for $m=n-1$,
$$
  \left(1-\frac{h_{n+1}}{h_n}\right)\left(1-\frac{h_n}{h_{n-1}}\right)=-\pp{}{t_1}\ln h_{n}\pp{}{s_1}\ln h_n.
$$

\subsection{Non-intersecting Brownian motions}
Consider $N$ non-intersecting Brownian motions $x_1(t),\ldots,x_N(t)$ in $\R$, leaving from distinct points
$\a_1<\ldots<\a_N$ and forced to end up at distinct points $\b_1<\ldots<\b_N$.  From the
Karlin-McGregor formula (see \cite{KM}), the probability that all $x_i(t)$ belong to $E\subset \R$ can be expressed
in terms of the Gaussian $p(t,x,y)=e^{-(x-y)^2/2 t }/\sqrt{2\pi t }$, as follows ($0<t<1$)
\begin{eqnarray}\label{Karlin1}
 \lefteqn{\BP_\a^{\b} \left(\mbox{all~$x_i(t)\in E $}\right)}\nonumber\\
 &:=&\BP_\a^{\b} \left(\mbox{all~$x_i(t)\in E $}
    \left|
      \begin{tabular}{c}
          $(x_1(0),\ldots,x_N(0)) =(\a_1,\ldots,\a_N)$\\
          $(x_1(1),\ldots,x_N(1)) =(\b_1,\ldots,\b_N)$\\
      \end{tabular}
    \right.\right)\nonumber\\
 &=&\frac1{Z_N}\int_{E^N}\det[p(t,\a_i,x_j)]_{1\leq i,j\leq N}\det[p(1-t,x_i,\b_j)]_{1\leq i,j\leq N}\prod_{i=1}^N dx_i\nonumber\\
 &=&\frac{1}{Z'_N}\int_{E^N}\prod_{i=1}^N e^{\frac{-x_i^2}{2t(1-t)}} dx_i
     \det\left[e^{\frac{\a_ix_j}{t}}\right]_{1\leq i,j\leq N}
     \det\left[e^{\frac{\b_ix_j}{1-t}}\right]_{1\leq i,j\leq N}
\end{eqnarray}
The limiting case where several points $\a$ and $\b$ coincide has been the object of many interesting
studies. It is obtained by taking appropriate limits of the formulae above. Just to fix the notation, consider
\begin{eqnarray*}
  (\a_1,\dots,\a_N)&=&(\overbrace{a_1,a_1,\dots,a_1}^{m_1},\overbrace{a_2,a_2,\dots,a_2}^{m_2},
             \dots,\overbrace{a_q,a_q,\dots,a_q}^{m_q})\\
  (\b_1,\dots,\b_N)&=&(\overbrace{b_1,b_1,\dots,b_1}^{n_1},\overbrace{b_2,b_2,\dots,b_2}^{n_2},\dots,
             \overbrace{b_p,b_p,\dots,b_p}^{n_p}),
\end{eqnarray*}
where $\sum_{\a=1}^q{a_\a}=\sum_{\b=1}^p{b_\b}=0$ and
\begin{equation*}
  a_1<a_2<\cdots a_q,\quad b_1<b_2<\dots<b_p,\quad\sum_{\a=1}^qm_\a=\sum_{\b=1}^pn_\b= N.
\end{equation*}%
Then, take the limit of (\ref{Karlin1}), make a change of variables in the second equality, use the standard matrix
identity in the third equality
$$
  \sum_{\sigma \in S_n}\det \left(a_{i,\sigma(j)}~b_{j,\sigma(j)}\right)_{1\leq i,j\leq n}
  =\det \left( a_{ik}\right)_{1\leq i,k\leq n}\det\left( b_{ik}\right)_{1\leq i,k\leq n},
$$
and distribute the integral and the Gaussian over the different columns; this yields
\begin{eqnarray*}
  \lefteqn{\BP_{\a}^{\b} \left( \mbox{all~$x_i(t)\in E $} \right) }\\
  &=&\frac{1}{Z_N}\int_{E^N}\prod_{i=1}^Ne^{-\frac{ x_i^2}{2 t (1- t )}} dx_i\\
  &&\times\det\left(\begin{array}{c}\left(x_j^{i }e^{\frac{a_1x_j}{t}}\right)_{\tiny
    \begin{array}{c}
      0\leq i<m_1\\
      1\leq j\leq N
    \end{array}}\\
  \\
  \vdots\\
  \\
  \left(x_j^{i }e^{\frac{a_qx_j}{t}}\right)_{\tiny
    \begin{array}{c}
       0\leq i<m_q\\
       1\leq j\leq N
    \end{array}}
  \end{array}\right)\cdot
  \det \left(
    \begin{array}{c}
      \left(x_j^{i }e^{\frac{b_1x_j}{1-t}}\right)_{\tiny
       \begin{array}{c}
          0\leq i<n_1\\
          1\leq j\leq N
       \end{array}}\\
  \\
  \vdots\\
  \\
  \left(x_j^{i }e^{\frac{b_qx_j}{1-t}}\right)_{\tiny\begin{array}{c}
                    0\leq i<n_p\\
                    1\leq j\leq N
                    \end{array}}
\end{array}\right)\\
&=&
   \frac{1}{Z'_N}\int_{  \tilde{E} ^N}
  \prod_{i=1}^Ne^{-\frac{ y_i^2}{2  }} dy_i
%
\\
&& \left.\times\det\left(\begin{array}{c}
  \left(y_j^{i }e^{ {\tilde a_1y_j}{ }}\right)_{\tiny\begin{array}{c}
                    0\leq i<m_1\\
                    1\leq j\leq N
                    \end{array}}\\
  \\
  \vdots\\ \\
  \left(y_j^{i }e^{ {\tilde a_qy_j}{ }}\right)_{\tiny\begin{array}{c}
                    0\leq i<m_q\\
                    1\leq j\leq N
                    \end{array}}
  \end{array}\right)\cdot
  \det \left(\begin{array}{c}
  \left(y_j^{i }e^{ {\tilde b_1y_j} }\right)_{\tiny\begin{array}{c}
                    0\leq i<n_1\\
                    1\leq j\leq N
                    \end{array}}\\
  \\
  \vdots\\ \\
  \left(y_j^{i }e^{ {\tilde b_qy_j}{ }}\right)_{\tiny\begin{array}{c}
                    0\leq i<n_p\\
                    1\leq j\leq N
                    \end{array}}
\end{array}\right)\right|
 _{\begin{array}{c}\tilde
 E=\frac{E}{\sqrt{t(1-t)}}
    \\
    \tilde a_i=\sqrt{\frac{1-t}{t}}a_i
    \\
    \tilde b_i=\sqrt{\frac{t}{1-t}}b_i
    \end{array}
}
 \\
 &=&\frac{N!}{Z'_N}\det\left( \left(\int_{\tilde E} dy~e^{-\frac{y^2}{2}}y^{i+j}e^{(\tilde
 a_{\a}+\tilde b_{\b})y}\right)
  _{\tiny\begin{array}{c}0\leq i< m_{\a}\\
                    0\leq j< n_{\b}
       \end{array}}
       \right)
       _{\tiny\begin{array}{c}1\leq \a\leq q\\
                    1\leq \b\leq p
       \end{array}}
  .\end{eqnarray*}
The numerator of this probability has exactly the form (\ref{intro1}) evaluated at $s_{\a}=t_{\b}=0$, with
the inner product given by (\ref{example})
$$
 \inn{x^i\psi_\a(x)}{y^j\varphi_{\b}(y)}
 =\int_{\tilde E} dy~e^{-\frac{y^2}{2}}y^{i+j}e^{(\tilde
 a_{\a}+\tilde b_{\b})y},
$$
upon setting $\psi_\a(x)=e^{\tilde a_{\a}x}$, $\varphi_\b(y)=e^{\tilde b_{\b}y}$ and
$d\mu(x,y)=\delta(x-y)e^{-y^2/2}\chi_{\tilde E}(y)\,dx$.  By multiplying each of the exponentials $e^{\tilde
a_{\a}y}$ and $e^{\tilde b_{\b}y}$ by $e^{-\sum_1^{\iy}s_{\a,k}y^k} $ and $e^{\sum_1^{\iy}
t_{\b,k}y^k} $ respectively, it follows that both the numerator and the denominator of the probability above,
\begin{eqnarray*}
  \lefteqn{\tau_{mn}(t_1,\dots,t_p;s_1,\ldots,s_q)}\\
   &=&\det\left(\left(\int_{\tilde E} dy~e^{-\frac{y^2}{2}}y^{i+j}e^{(\tilde a_{\a}+\tilde b_{\b})y
      +\sum_1^{\iy}(t_{\b,k}-s_{\a,k})y^k}\right)_{\tiny
      \begin{array}{c}
         0\leq i< m_{\a}\\
         0\leq j< n_{\b}
       \end{array}}
       \right)_{\tiny
      \begin{array}{c}
         1\leq \a\leq q\\
         1\leq \b\leq p
      \end{array}}       \\
      &=&
       \det\left(\left(\inn{x^i\psi^{-s}_\a(x)}{y^j\varphi^t_{\b}(y)}
                  \right)_{\tiny\begin{array}{c}0\leq i< m_{\a}\\
                                                   0\leq j< n_{\b}
                                  \end{array}}
           \right)_{\tiny\begin{array}{c}1\leq \a\leq q\\
                                            1\leq \b\leq p
                           \end{array}}
\end{eqnarray*}
and the same expression for $E=\R$, satisfy the bilinear identity for $p+q$-component KP and, in particular, all
the general relations and identities, mentioned in this paper, namely (\ref{introa}) and (\ref{eq:PDE_in_prop12}),
(\ref{eq:PDE_in_prop13}), (\ref{eq:PDE_in_prop14}). Note the equations are independent of the set $E$.

\smallskip
\emph{In particular}, for $n$ non-intersecting Brownian motions, departing from the origin, with $n_1$ paths forced
to end up at $-a$ and $n_2$ paths forced to end up at $a$, we have for $0< t< 1$,
\begin{equation*}
  \BP_0^{\pm a}(\mbox{all}~x(t)\in \tilde E) =
    \frac{1}{Z_n} \det\left(
       \begin{array}{l}
          (\mu_{ij}^+)_{0\leq i \leq n_1-1,~0\leq j\leq n_1+n_2-1 }\\  \\
          (\mu_{ij}^-)_{0\leq i \leq n_2-1,~0\leq j\leq n_1+n_2-1 }\\
       \end{array}
     \right)
\end{equation*}
where
\begin{equation}\label{moment1}
  \mu_{ij}^{\pm}:=\int_{E} x^{i+j}e^{-\frac{x^2}{2}\pm\a x} dx,
\end{equation}%
with the change of variables
$$
  \a=a\sqrt{\frac{2t}{1-t}}\mbox{~~and~~} E=\tilde E\sqrt{\frac{2 }{t(1-t)}}.
$$
In a similar way, for several times $0=t_0<t_1<\ldots<t_m<t_{m+1}= 1$,
$$
  {\BP_0^{\pm a}(\mbox{all~}x_i(t_1)\in\tilde E_1,\ldots,\mbox{all~}x_i(t_m)\in\tilde E_m)}
  =
  \frac{1}{Z_n} \det\left(
    \begin{array}{l}
       (\mu_{ij}^+)_{0\leq i \leq n_1-1,~0\leq j\leq n_1+n_2 -1}\\  \\
       (\mu_{ij}^-)_{0\leq i \leq n_2-1,~0\leq j\leq n_1+n_2 -1}\\
    \end{array}
  \right),
$$
where
\begin{equation}\label{moment2}
   \mu_{ij}^{\pm}=
   \int_{\prod_1^m E_k}(x_{1})^{j-1}(x_{m})^{i-1}
      e^{-\frac{1}{2}\sum_{\ell=1}^{m}x_{\ell}^2\pm\a x_m+ \sum_{\ell=1}^{m-1}c_\ell x_\ell x_{\ell+1}}
      \prod_{\ell=1}^m dx_\ell,
\end{equation}%
with the change of variables
$$
  \a=a\sqrt{\frac{2(t_{m}-t_{m-1})}{(1-t_m)(1-t_{m-1})}},\quad
  E_\ell=\tilde E_\ell\sqrt{\frac{2(t_{\ell+1}-t_{\ell-1})}{(t_{\ell+1}-t_{\ell})(t_{\ell}-t_{\ell-1})}},
$$
and
$$
  c_j =\sqrt{ \frac{(t_{j+2}-t_{j+1})(t_{j}-t_{j-1})}{(t_{j+2}-t_{j})(t_{j+1}-t_{j-1})}},~~\mbox{for}~1\leq j\leq m-1.
$$
We now introduce the inner products
$$
  \inn{f}{g}_1= \int_E f(x)g(x)F_1(x)dx\mbox{~~with } ~ F_1(x)=e^{-\frac{x^2}{2}}
$$
and ($m\geq 2 $)
$$
  \inn{f}{g}_m=\int_{\prod _1^m E_k} f(x_1)g(x_m)F_m(x_1,\ldots,x_m)dx_1\ldots dx_m,
$$
with
$$
  F_m(x_1,\ldots,x_m):=\left(\prod_1^m e^{-\frac{x_{\ell}^2}{2}}\right)e^{\sum_{ p,q\geq 1}\sum_{\ell=1}^{m-1}
    c^{(\ell)}_{pq} x_{ \ell } ^px_{\ell+1}^q+\sum_{{\ell}=2}^{m-1}\sum_{r=1}^{\iy}\gamma^{(\ell)}_rx_{ {\ell} } ^r}.
$$
The precise form of $F_m$ does not matter very much for the purpose of this paper, but does play a crucial role in
satisfying the Virasoro constraints.

\smallskip

In these two sets of moments (\ref{moment1}) and (\ref{moment2}), we insert extra time-parameters, as follows,
which can then be identified with the moments appearing in (\ref{intro1}),
\begin{eqnarray*}
  \mu^\pm_{ij}( s,u,v)
    &=&\ds{\int_{E}}x^{i+j}e^{-\frac{x^2}{2}\pm ax\pm\b x^2}e^{\sum_1^{\iy}(s_k-\left({{u_k}\atop{v_k}}\right))x^k}dx\\
    &=&\inn{x^ie^{-\sum_1^{\iy} s_kx^k}}{x^je^{\sum_1^{\iy}\left({{u_k}\atop{v_k}}\right)x^k}e^{\pm\a x \pm \b
    x^2}}_1,
\end{eqnarray*}
and
\begin{eqnarray*}
  \mu^\pm_{ij}( s,u,v)
    &=&\int_{\prod_1^m E_k}x_{1}^{i} x_{m}^{j}F_m(x_1,\ldots,x_m)e^{\sum_{k=1}^{\iy}\left(s_kx_{1}^k-\left({{u_k}\atop{v_k}}\right)
      x_{m}^k\right)}\prod_{\ell=1}^m dx_{\ell}\\ \\
    &=&\inn{x^ie^{-\sum_1^{\iy}s_kx^k}}{x^je^{ \sum_1^{\iy}\left({{u_k}\atop{v_k}}\right)x^k}e^{\pm\a x\pm\b x^2}}_m
\end{eqnarray*}
In both cases, we have $q=1$ and $p=2$, and $m:=n_1+n_2$, leading to the introduction of three sets of times
$s_i:=-s_{1i}$, $u_i:=-t_{1i}$ and $v_i:=-t_{2i}$.Thus, from the general theory, the numerator of both
probabilities,
$$
  \tau_{n_1,n_2}=\det\left(
    \begin{array}{l}
      (\mu_{ij}^+)_{0\leq i \leq n_1-1,~0\leq j\leq n_1+n_2 -1}\\  \\
      (\mu_{ij}^-)_{0\leq i \leq n_2-1,~0\leq j\leq n_1+n_2 -1}\\
    \end{array}
  \right)
$$
satisfies the bilinear identity for the 3-component KP and, in particular, the PDE's and thus $\tau_{n_1,n_2}$
satisfies the single PDE
\begin{eqnarray*}
\frac{\p}{\p s_1}\ln \frac{\tau_{n_1 +1,n_2 }}
                           {\tau_{n_1 -1,n_2 }}
                           &=&\frac
      {\frac{\p^2}{\p s_2\p u_1}\ln \tau_{n_1 ,n_2 }}
      {\frac{\p^2}{\p s_1\p u_1}\ln \tau_{n_1 ,n_2 }}
  \label{12}\\
    -\frac{\p}{\p u_1}\ln \frac{\tau_{n_1 +1,n_2}}
                          {\tau_{n_1 -1,n_2 }}
                          & =&     \frac
      {\frac{\p^2}{\p s_1\p u_2}\ln \tau_{n_1 ,n_2 }}
      {\frac{\p^2}{\p s_1\p u_1}\ln \tau_{n _1,n_2 }}.
\end{eqnarray*}
$$
  \frac{\p}{\p u_1}\frac {\frac{\p^2}{\p s_2\p u_1}\ln \tau_{n_1 ,n_2 }} {\frac{\p^2}{\p s_1\p u_1}\ln\tau_{n_1,n_2 }}
  + \frac{\p}{\p s_1} \frac {\frac{\p^2}{\p s_1\p u_2}\ln \tau_{n_1 ,n_2 }} {\frac{\p^2}{\p s_1\p u_1}\ln\tau_{n_1,n_2 }} =0
$$
and the same PDE with $u_i$ replaced by $v_i$. These PDE's play a crucial role in establishing the PDE for the
Pearcey process; see \cite{AMPearcey}.

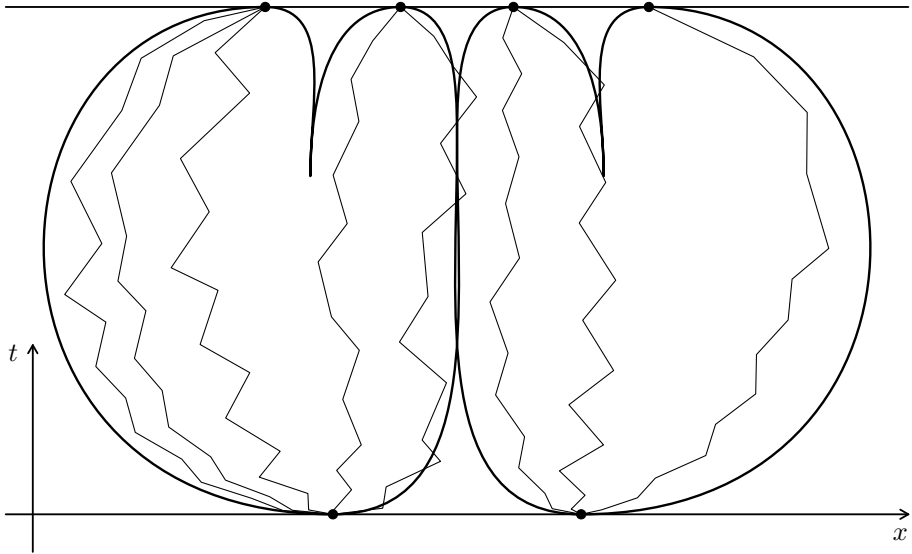
$$\vcenter{\hbox{\epsfbox{pictures.100}}}$$

The methods developed in this paper should enable one to study more complicated situations of non-intersecting
Brownian motions, as indicated in the figure above. The curves in the $(x,t)$-plane are the boundary of the
equilibrium measure as a function of time. When two curves meet, one expects to see a new infinite-dimensional
diffusion in that neighborhood, beyond the Pearcey process.

%

\def\cydot{\leavevmode\raise.4ex\hbox{.}}

\end{document}